\documentclass[]{iopart}
\usepackage[T1]{fontenc}
\usepackage[utf8]{inputenc}
\usepackage{url}
\expandafter\let\csname equation*\endcsname\relax
\expandafter\let\csname endequation*\endcsname\relax 
\usepackage{amsmath}
\usepackage{amsthm}
\usepackage{amssymb}
\usepackage{graphicx}
\usepackage{setspace}
\usepackage[czech,english]{babel}
\usepackage{mathrsfs}
\usepackage{mathtools}
\usepackage{physics}
\usepackage{pgfplots}
\usepackage{enumitem}
\usepackage{float}
\usepackage{appendix}
\usepackage{tensor}
\usepackage{tocloft}
\usepackage{mathtools}
\usepackage[caption = false]{subfig}

\renewcommand{\a}{\alpha}
\renewcommand{\b}{\beta}
\newcommand{\g}{\gamma}
\renewcommand{\d}{\delta}

\newcommand{\fii}{\varphi}
\renewcommand{\e}{e^{i\varphi}}
\newcommand{\ee}{e^{-i\varphi}}

\newtheorem{theorem}{Theorem}[section]
\theoremstyle{definition}
\newtheorem{definition}[theorem]{Definiton}
\newtheorem{remark}[theorem]{Remark}
\newcommand{\ad}[1]{#1^{\dagger}}
\newcommand{\dg}[1]{#1^{\dagger}}
\renewcommand{\H}{\mathscr{H}}

\newcommand{\h}{\mathscr{H}}
\newcommand{\bh}{\mathcal{B}\left(\mathscr{H}\right)}
\DeclareMathOperator{\Ker}{Ker}
\renewcommand{\a}{\alpha}
\renewcommand{\b}{\beta}
\renewcommand{\d}{\delta}

\pgfplotsset{compat=1.17}

\begin{document}

\title[Asymptotic phase-locking and synchronization in two-qubit systems]{Asymptotic phase-locking and synchronization in two-qubit systems}

\author{D. Štěrba, J. Novotný, I. Jex}
\address{Department of Physics, FNSPE, Czech Technical University in Prague, Břehová 7, 115 19 Praha 1, Czech Republic}

\begin{abstract}
The paper concerns spontaneous asymptotic phase-locking and synchronization in two-qubit systems undergoing continuous Markovian evolution described by Lindbladian dynamics with normal Lindblad operators. Using analytic methods, all phase-locking-enforcing mechanisms within the given framework are obtained and classified. Detailed structures of their respective attractor spaces are provided and used to explore their properties from various perspectives. Amid phase-locking processes those additionally enforcing identical stationary parts of both qubits are identified, including as a special case the strictest form of synchronization conceivable. A prominent basis is presented which reveals that from a physical point of view two main types of phase-locking mechanisms exist. The ability to preserve information about the initial state is explored and an upper bound on the amplitude of oscillations of the resulting phase-locked dynamics is established. Permutation symmetry of both asymptotic states and phase-locking mechanisms is discussed. Lastly, the possibility of entanglement production playing the role of a phase-locking witness is rebutted by three analytically treatable examples.
\end{abstract}

\section{Introduction}
\label{sectionIntroduction}

Spontaneous synchronization was for the first time reportedly observed by Christiaan Huygens who took notice of the tendency of two pendulum clocks mounted on a common bar to adjust to anti-phase oscillations, and described the observation in his letters as early as in February 1665 \cite{HuygensOriginal}. Since then, synchronization has been thoroughly explored in a great variety of classical systems \cite{SynchBook} and recently the study of this ubiquitous phenomenon entered the quantum realm.

In both classical and quantum domain, synchronization is a very broad term. Various viewpoints, understandings and hence definitions of it have been introduced \cite{QSmeasures1}, \cite{QSmeasures2}, \cite{QSmeasures3}, \cite{ConsensusGuodong}, \cite{BruderMinimalSystem}, and systems investigated are numerous. The first works on the subject were typically concerned with forced synchronization induced by an external field, the so-called entrainment. A quantum system is synchronized with an external signal replacing its original dynamics completely. Examples include a driven oscillator \cite{FS1drivenoscillator}, an oscillator coupled to a qubit \cite{FS2qubitcoupledtooscillator} or systems of van der Pol oscillators \cite{FS3vanderPolOscillators}.
It was recently argued \cite{BruderMinimalSystem} that the smallest quantum system which can be synchronized in this way is spin $1$. The idea behind the paper is based on a system with a dissipator sending it towards a stationary single point limit, which is then deformed by an external driving force. That results in nontrivial time evolution, naturally phase-locked to the inducing signal. The reason to exclude qubit from consideration was a lack of what the authors regard as a proper limit cycle, which was opposed in \cite{BergliGeneralizationOfBruder} by introduction of the concept's generalization. Needless to say, the argument is not relevant to our work as what we consider here are systems with naturally non-stationary asymptotic limits, or limit cycles, exhibiting synchronized behaviour.
Another noteworthy direction of research is represented by synchronization protocols, proposals of how to exploit system properties such as entanglement to achieve clock synchronization between two parties \cite{SP1clocksnychpriorentanglement}.

A significant part of research activities on synchronization phenomena is focused on spontaneous synchronization, the situation when individual subsystems tune their local dynamics to a common pace due to the presence of coupling. Prevalent are studies of transient synchronization, also referred to as metastable synchronization, which results from a time-scale separation of decay rates of individual parts of an initial state in a dissipative system. The basic principle is well-described for example in \cite{TS1comprehensiveOscAndSpin}. In such a case the system goes through a long-lasting yet temporary phase of, to some degree, synchronized evolution, eventually approaching relaxation in the asymptotics. Among the examined systems are oscillator networks \cite{TS3oscillatorNetworks}, \cite{TS5twooscillators}, spin systems \cite{TS4spinsystems}, atomic lattices \cite{TS2atomiclattice}, qubits in bosonic environment \cite{TS6twoqubitssubradiance}, collision models \cite{Iskender}, simple few-body systems in dissipative environments \cite{Iskender2} and many other.

Aside from the study of transient synchronization, it was demonstrated that a decoherence process can be designed whose decoherence-free subspace \cite{DFS} permits solely synchronized evolution \cite{ConsensusGuodong,Buca}. Consequently, the system is driven towards an asymptotic regime with perfectly synchronized individual dynamics. In contrast to the transient synchronization, the synchronized part of evolution is not accompanied by relaxation and it survives for an infinitely long time. Such a process is referred to as spontaneous asymptotic synchronization. Naturally, all such asymptotically synchronizing mechanisms take effect only if the individual internal dynamics are detuned; once tuned, they leave the state of the system untouched.

In this work we study spontaneous asymptotic phase synchronization and, more generally, spontaneous asymptotic phase-locking, further in the text simply referred to as synchronization or phase-locking. We aim to better understand how a decoherence process must be designed in order to enforce a prescribed asymptotic phase difference between dynamics of individual subsystems. Thus, instead of analyzing synchronization properties of a particular quantum system with a given dynamics, we assume a bipartite qubit-qubit system with general but identical internal single-qubit dynamics. Within a broad class of decoherence mechanisms we identify all those capable of enforcing synchronization of the two parties. As presence of a third party, such as but not limited to an environment coupled to the subsystems, is essential for synchronization, the evolution of the studied system is conveniently modelled as a continuous quantum Markov process generated by a Lindbladian. In our approach, we search for possible candidates for synchronization mechanisms within a class of quantum Markov dynamical semigroups (QMDS) with normal Lindblad operators. The reason is twofold. Firstly, the asymptotics of a system undergoing evolution given by a QMDS can be analytically analyzed using attractor theory for quantum Markov processes \cite{JJ}. This theory relies on the presence and knowledge of a faithful invariant state, neither of which is generally guaranteed. However, QMDS with normal Lindblad operators are always unital and, consequently, all of them preserve the maximally mixed state. Moreover, this choice significantly simplifies the attractor equations determining system's asymptotic properties. The ability to address the problem analytically is a noteworthy advantage, especially compared to the studies of transient synchronization, which often rely on numerical treatment. Analytical approach enables us to work with a precise definition of synchronization independent of the choice of a particular synchronization measure. Secondly, QMDS with normal Lindblad operators constitute a large set of possible dynamics, including several families of QMDS with phase-locking-enforcing Lindblad operators for an arbitrary given phase. 

Having first identified all synchronization and phase-locking mechanisms expressible by a single Lindblad operator, our approach further allows us to study their properties. In particular, for each class of mechanisms we reveal the structure of the corresponding attractor spaces. Based on the results, we show how individual Lindblad operators can be combined to construct additional phase-locking mechanisms and thereby cover all possible phase-locking-enforcing evolutions within the framework and provide their classification. The structure of the attractor spaces also indicates the presence of a special phase-locking basis, in which it turns out that from a physical point of view there exist only two main types of two-qubit phase-locking mechanisms. Mechanisms of the first type are equipped with a decoherence-free subspace admitting only phase-locked two-qubit states. Mechanisms of the second type do not have a decoherence-free subspace at all. Their attractor spaces contain components with different phase delays, but the unwanted contributions to the reduced single-qubit states mutually cancel and vanish. With the insight into the synchronization mechanisms we proceed and study visibility of the resulting asymptotic reduced single-qubit dynamics, capability to preserve information about an initial state, global symmetries of the synchronized asymptotic states and their relation to the symmetries of the synchronization mechanisms, and entanglement generation during the evolution towards asymptotically synchronized states.

The paper is organized as follows. In chapter \ref{section_theory} we summarize the necessary theoretical background of QMDS and discuss particular aspects and tools used in the actual analysis of asymptotic synchronization and phase-locking. In chapter \ref{section_synchronization} we give suitable definitions of phase synchronization and phase-locking used throughout the work. In the following chapter \ref{section2QSynchronizationMechanisms} the two-qubit system is thoroughly examined. All phase-locking-enforcing Lindblad operators are found and classified. It is explained how they can be combined in a Lindbladian and those which additionally enforce identical stationary parts of the asymptotic states are identified. In chapter \ref{sectionAsymptoticProperties} the attractor spaces of all phase-locking mechanisms are presented and followed by a detailed description of the corresponding asymptotic dynamics. Based on the results, additional properties such as visibility of the resulting phase-locked single-qubit oscillations or entanglement generated by phase-locking mechanisms are discussed. Lastly, in chapter \ref{section_Conclusions} the obtained results are briefly summarized. Technical details and lengthy calculations are left for \ref{AppendixDerivation} and used parameterization of normal two-qubit operators is given in \ref{AppendixNormalMatrices}.

\section{Quantum Markov Dynamical Semigroups and Asymptotics of Lindbladian Dynamics}
\label{section_theory}

Synchronization of individual dynamics is expected to arise in systems weakly coupled via a common environment. This scenario is well described and frequently studied using trace-preserving continuous quantum Markov processes (CQMP). Dynamics of states, density operators $\rho$ on a finite-dimensional Hilbert space $\h$, is given by a master equation in the well-known Lindblad (also Gorini–Kossakowski–Sudarshan–Lindblad or GKSL) form
\begin{equation}
\label{Lindblad}
\frac{d\rho}{dt}=\mathcal{L}(\rho) = -i\comm{H}{\rho} + \sum_j L_j \rho \dg{L}_j - \frac{1}{2}\acomm{\dg{L}_j L_j}{\rho},
\end{equation}
where $H$ is the Hamiltonian of the system and $L_j$ are referred to as Lindblad operators. The generator $\mathcal{L}$, called Lindbladian, acts on the space of all operators $\bh$ on the Hilbert space $\h$. Here, the commutator with the Hamiltonian corresponds to the unitary part of the evolution, akin to a closed system, whereas the terms with Lindblad operators account for the effects of the surrounding environment and possible environment-assisted interactions of the involved parties.
The resulting evolution of states $\rho(t)=\mathcal{T}_t(\rho(s))$ governed by a quantum dynamical semigroup (QMDS) $\mathcal{T}_t=\exp(t\mathcal{L})$ is typically much harder to analyze compared to that of closed quantum systems. This is due to the complexity of the, in general, non-normal generator $\mathcal{L}$, which need not even be diagonalizable. However, with the aim to study asymptotic synchronization, we focus solely on the asymptotic regime of CQMP whose analytical treatment was developed in \cite{JJ, Albert}.

Let us briefly summarize its main points relevant for this work. The asymptotic spectrum $\sigma_{as}(\mathcal{L})$ of a generator $\mathcal{L}$ is the set of all purely imaginary points of its spectrum $\sigma(\mathcal{L})$
\begin{equation}
\sigma_{as}(\mathcal{L}) = \{\lambda \in \sigma(\mathcal{L}), \mathrm{Re} \lambda = 0\}.
\end{equation}
The subspace spanned by the eigenvectors of $\mathcal{L}$ associated with eigenvalues from the asymptotic spectrum is called the attractor space $\mathrm{Att}(\mathcal{T})$ of the QMDS $\mathcal{T} = \exp (t\mathcal{L})$,
\begin{equation}
\mathrm{Att}(\mathcal{T}) = \bigoplus_{\lambda \in \sigma_{as}(\mathcal{L})} \Ker(\mathcal{L}-\lambda I).
\end{equation}
We commonly refer to elements $X \in \mathrm{Att}(\mathcal{T})$ as attractors. It can be shown that all asymptotic states of CQMP, including steady states, belong to this subspace, and thus the asymptotic dynamics of CQMP can always be diagonalized. Provided that we are able to construct a basis  $\{X_{\lambda , i}\}$ of $\mathrm{Att}(\mathcal{T})$, where $\lambda$ goes through all eigenvalues from the asymptotic spectrum and $i$ refers to their possible degeneracies, and its dual basis $\{X^{\lambda,i}\}$ satisfying $\Tr{\dg{X}_{\lambda,i}X^{\lambda',j}} = \delta_{\lambda \lambda'}\delta_{ij}$, we can express the asymptotic dynamics of an arbitrary initial state $\rho(0)$ as 
\begin{equation}
\label{AsymptoticDynamics}
\rho_{as}(t) = \sum_{\lambda \in \sigma_{as}(\mathcal{L}), i} e^{\lambda t}\Tr\left\{\left(X^{\lambda , i}\right)^{\dagger}\rho(0)\right\} X_{\lambda , i},
\end{equation}
which satisfies the limit
\begin{equation}
\lim_{t \to \infty}	\norm{\rho(t) - \rho_{as}(t)} = 0.
\end{equation}
In order to construct both basis we typically need a so-called maximal stationary state, i.e. an invariant state of the CQMP with maximal rank. Such a state always exists, it is not unique in general, and any asymptotic state is restricted to its support. Unfortunately, different CQMP may, naturally, have different maximal stationary states, and those are not always easily obtained, which makes analytical treatment of asymptotic dynamics of general CQMP virtually impossible.

To avoid this obstacle, we restrict ourselves to normal Lindblad operators $L_j$ in the generator \eqref{Lindblad}. There are two main reasons for the choice. Firstly, normal Lindblad operators generate QMDS which are all unital, i.e. the maximal mixed state is always their maximal invariant state due to 
\begin{equation}
\mathcal{L}(I) = \sum_j L_j \dg{L}_j - \dg{L}_j L_j = 0,
\end{equation}
which is satisfied for an arbitrary number and combination of Lindblad operators, in particular for a single Lindblad operator, if and only if $\comm{L_j}{\dg{L}_j} = 0$, that is if the operators $L_j$ are normal. Consequently, they constitute a set of possible Lindblad operators for which the analysis of asymptotic dynamics of corresponding QMDS is significantly easier. 

Indeed, the eigenspaces $\Ker(\mathcal{L}-\lambda I)$, $\lambda \in \sigma_{as}(\mathcal{L}),$ forming the attractor space $\mathrm{Att}(\mathcal{T}$) of a unital QMDS are mutually orthogonal with respect to the Hilbert-Schmidt scalar product $(A,B)=\Tr{A^{\dagger}B}$, and the same holds for the $\mathrm{Att}(\mathcal{T})$ and its complement $Y$, which accounts for the asymptotically vanishing part of the evolution. Additionally, an orthogonal basis of the attractor space can be chosen such that the elements of the dual basis $X^{\lambda,i}$ can be identified with $X_{\lambda,i}$, and this basis can be procured using a simplified version of the structure theorem \cite{JJ}.

\begin{theorem}
	\label{StructureTheorem}
	Let $\mathcal{T}_t : \bh \to \bh$ be a trace-preserving QMDS with generator $\mathcal{L}$ of the form \eqref{Lindblad} where all Lindblad operators $L_j$ are normal. An element $X \in \bh$ is an attractor of $\mathcal{T}_t$ associated with eigenvalue $\lambda \in \sigma_{as}(\mathcal{L})$ if and only if it holds
	\begin{equation}
	\label{StructureTheoremEq1}
	\comm{L_j}{X} = \comm{\dg{L}_j}{X} = 0,
	\end{equation}
	\begin{equation}
	\label{StructureTheoremEq3}
	\comm{H}{X} = i\lambda X.
	\end{equation}
\end{theorem}
The theorem reveals how the structure of the generator $\mathcal{L}$ is closely intertwined with the attractor space of the corresponding QMDS and represents a key tool in our analysis of two-qubit synchronization and phase-locking mechanisms performed in this work.

Secondly, the evolution of a unital QMDS within its attractor space, i.e. the asymptotic evolution, is unitary, generated by the system Hamiltonian $H$. Therefore, asymptotic states of unital QMDS evolve without the influence of environment represented by the Lindblad operators $L_j$. This is a key property which ensures that in the absence of direct interactions, when all interactions are assumed to be environment-mediated and hence represented by the Lindblad operators, the evolution of asymptotically synchronized subsystems of a composite quantum system is driven solely by their local free Hamiltonians. For a non-unital QMDS this is not generally true \cite{JJ}.

\section{Definitions of synchronization and phase-locking in CQMP}
\label{section_synchronization}

Attractor theory introduced in the previous section represents a well-suited analytic tool for ascertaining whether a given two-partite quantum system undergoing CQMP evolves towards an asymptotic trajectory with a desired mutual phase relationship between the individual parties. Conversely, it enables us to construct systems with predetermined asymptotic properties, in particular systems which induce asymptotic synchronization or phase-locking independently of initial conditions.
Due to the ability to approach the problem analytically and considering the existence of a well-defined asymptotic limit for all initial conditions, we do not have to rely on a particular choice of a synchronization measure, as it is common in the literature, which often relies on numerical treatment.
Instead, we provide and employ exact definitions of synchronized and phase-locked subsystem evolution of bipartite quantum systems. It contributes to clarity and unambiguity of the results, and is all the more desirable since our main goal is to find CQMP achieving perfect synchronization. 

To proceed, assume a single quantum system with an associated Hilbert space $\H$ and Hamiltonian $H_0$. A bipartite system of identical subsystems, copies of our original quantum system, is then associated with the Hilbert space $\h^{\otimes 2} = \h_A \otimes \h_B$, $\h_A = \h_B = \h$. The whole bipartite system is supposed to undergo a CQMP given by \eqref{Lindblad} with a free Hamiltonian $H=H_A+H_B = I \otimes H_0 + H_0 \otimes I$. While the internal dynamics of both subsystems is driven by the same local Hamiltonian, their initial conditions differ in general and neither need their initial state to be separable. The Linbladian part comprises the influence of a common environment and environment-assisted interactions. Let us denote  $\rho_A(t) = \Tr_{B}\rho(t)$, $\rho_B(t) = \Tr_{A}\rho(t)$ the reduced states of the global state $\rho(t) \in \mathcal{B}(\h^{\otimes 2})$, obtained as a partial trace of the state $\rho(t)$ of the bipartite system over the remaining subsystem. In the following, we define several mutually related concepts concerning asymptotic synchronization and phase-locking which we intend to investigate. The first definition of synchronized subsystems requires that apart from their stationary parts both subsystems evolve identically.

\begin{definition}
	\label{DefSynchronization}
	We say that two identical subsystems $A$ and $B$ of a bipartite quantum system in a state $\rho(t) \in \mathcal{B}(\h^{\otimes 2})$ are synchronized if there exists a stationary operator $\rho_{const}\in \bh$ such that the reduced states $\rho_A (t), \rho_B (t)$ satisfy
	\begin{equation}
	\label{defSynchEq}
	\rho_A (t) - \rho_B( t) = \rho_{const},  \quad \forall t.
	\end{equation}

Obviously, this stationary part is represented by a hermitian operator with zero trace.
The subsystems achieve asymptotic synchronization if they become synchronized in the limit $t \to \infty$, i.e.
	\begin{equation}
	\label{AsymptoticSynch}
	\lim_{t \to \infty} \norm{\rho_A (t) - \rho_B (t) - \rho_{const}} = 0.
	\end{equation}	
	
Naturally, we refer to the states of synchronized systems as synchronized, too. In this context, we also call the global evolution with synchronized subsystems synchronized, and say that a bipartite system achieves asymptotic synchronization if its subsystems do. We say that a quantum dynamical semigroup $\mathcal{T}_t$, and for a given internal dynamics also its generating Lindblad operators $\{L_j\}$ synchronize, enforce synchronization, constitute a synchronization mechanism or simply are synchronizing if asymptotic synchronization is achieved for an arbitrary initial state $\rho (0)$. 
\end{definition}

According to this definition, the subsystems are synchronized if the non-stationary parts of their states are equal and undergo the same internal evolution, which in the asymptotic limit is the unitary evolution given by the internal Hamiltonian $H_0$. Analogously, synchronization can be defined via synchronization of local observables in the Heisenberg picture, or via synchronization of expectation values. In fact, the definition could be rewritten as the expectation values of all local observables being synchronized up to a constant. In the freely evolving asymptotics, for the low-dimensional case of qubits our definition is even equivalent to synchronization of a single local observable given by the Pauli matrix $\sigma_x$, respectively to synchronization of its expectation values $\expval{\sigma_{x A}}(t) = \expval{\sigma_{x B}}(t)$ for any initial state.

We intentionally allow a difference of a stationary operator between the synchronized states, imposing constraints only on the dynamical part. To distinguish the situation where the entire reduced subsystem states are the same, we introduce a second, more restrictive definition. 

\begin{definition}
	\label{DefCompleteSynchro}
	We speak of complete synchronization of subsystems $A$ and $B$ if the reduced states $\rho_A (t), \rho_B (t)$ are identical\footnote{This can equivalently be viewed as synchronization of all observables.}, i.e.
\begin{equation}
	\label{defCompleteSynchEq}
	\rho_A (t) = \rho_B( t),  \quad \forall t,
	\end{equation}
	and of asymptotic complete synchronization if
	\begin{equation}
	\lim_{t \to \infty} \norm{\rho_A (t) - \rho_B (t)} = 0.
	\end{equation}
\end{definition}

Let us note that recently similar concepts of synchronization were adopted in \cite{Buca}. Understanding of synchronization, nonetheless, varies considerably across literature.

A straightforward generalization of synchronization is phase-locking, a process of establishing a given constant phase shift between otherwise identical internal dynamics of two subsystems. Similarly to synchronization, we shall distinguish phase-locking of subsystems with in general different stationary parts and of those with coinciding stationary parts. 
\begin{definition}
	\label{DefPhaseLocking}
Let $\rho_{A,st}$ and $\rho_{B,st}$ be the maximal stationary parts of subsystem states $\rho_A(t)$ and $\rho_B(t)$, and correspondingly let $\rho_{A,dyn}(t)$ and $\rho_{B,dyn}(t)$ be the remaining dynamical time-evolving parts, i.e.
\begin{equation}
\rho_{A/B}(t) = \rho_{A/B,st} + \rho_{A/B,dyn}(t). 
\end{equation}
In view of relation \eqref{AsymptoticDynamics}, both stationary parts are always nonzero and uniquely defined in the asymptotics, as they are directly linked to the attractors associated with eigenvalue $\lambda=0$. In particular, they are determined by the partial traces of the linear combination of projections of an initial state onto said attractors. 

Using this decomposition, we say that the subsystems A and B, in this order, are phase-locked with a phase shift $\varphi \in [0,2\pi)$ if the dynamical parts of their states satisfy
	\begin{equation}
	\label{PhaseLockingAB}	
	\rho_{A,dyn} (t) = e^{i\varphi} \rho_{B,dyn} (t),
	\end{equation}
i.e. if they differ exactly by a phase factor $e^{i\varphi}$. Asymptotic phase-locking is achieved when
	\begin{equation}
	\lim_{t \to \infty} \norm{\rho_{A,dyn} (t) - e^{i\varphi} \rho_{B,dyn} (t)} = 0.
	\end{equation}
For a phase shift $\varphi = 0$ phase-locking reduces to synchronization. Therefore, as an equivalent to the expression phase-locking we hereby introduce and hereafter parallelly use the term generalized synchronization.
\end{definition}

\begin{definition}
We speak of phase-locking with a phase shift $\varphi \in [0,2\pi)$ and simultaneous synchronization of the stationary parts, if in addition to \eqref{PhaseLockingAB} it holds
	\begin{equation}
	\label{ComplSynchAB}
	\rho_{A,st} = \rho_{B,st},
	\end{equation}
and we call it generalized complete synchronization. The remaining terminology is defined in a similar fashion.
\end{definition}

\begin{remark}
The definitions of synchronization and phase-locking are trivially satisfied by stationary states. In particular, CQMP having solely attractors associated with eigenvalue $\lambda=0$ always achieve asymptotic (generalized) synchronization. There also exist CQMP such that the overall evolution of the composite system is asymptotically non-stationary but only stationary attractors contribute to the reduced subsystem dynamics. In other words, such processes gradually destroy local evolution. We would call it trivial synchronization, in different context similar concepts are called consensus formation \cite{ConsensusGuodong}. However, since our prime goal is to find QMDSs enforcing synchronization or phase-locking of nontrivial internal subsystem dynamics, we are not interested in such situations in this paper.
\end{remark}

\section{Two-qubit synchronization and phase-locking}
\label{section2QSynchronizationMechanisms}

In this part, we investigate possible mechanisms of generalized synchronization, i.e. synchronization and phase-locking, for a system of two qubits. Our primary aim is to construct all possible Lindbladians with normal Lindblad operators such that the two-qubit evolution given by the corresponding QMDS inevitably results in their synchronization, respectively phase-locking. For clarity of presentation, we explain our analytical approach and provide the final classification of all two-qubit generalized synchronization mechanisms, while technical details and lengthy calculations are left for the \ref{sec:AttractorSpaces}. \\

Let $\h_0$ be the Hilbert space corresponding to a single qubit and $H_0 = \mbox{diag}(E_0,E_1)$ be the Hamiltonian in the basis of its eigenvectors $\ket{0}$ and $\ket{1}$. A system of two identical qubits is then associated with the Hilbert space $\h = \h_0 \otimes \h_0$ and a free Hamiltonian $H = H_0 \otimes I + I \otimes H_0$. We will work with a free Hamiltonian accounting for the internal evolution of qubits, which ensures that both subsystems have the same local periodic dynamics with identical intrinsic frequency $\omega = E_0 - E_1 = \Delta E$, and consider all mutual interactions to be mediated by the environment and described by the Lindblad operators. Let $\mathscr{B} = (\ket{00},\ket{01},\ket{10},\ket{11})$ be a computational basis of $\h$, using the standard notation $\ket{ij} = \ket{i} \otimes \ket{j}$, to which we stick throughout the entire paper, unless specified otherwise.

\subsection{General approach to identifying phase-locking mechanisms}

To begin with, we study the case of just a single normal Lindblad operator $L$ in the Lindbladian. Once we have a complete solution to this problem, generalization to an arbitrary finite number of normal Lindblad operators is straightforward. Therefore, in the following we assume a form of the generator \eqref{Lindblad} with exactly one normal Lindblad operator, namely

\begin{equation}
\label{LindbladOneL}
\mathcal{L}(\rho) = -i\comm{H}{\rho} +  L \rho \dg{L} - \frac{1}{2}\acomm{\dg{L} L}{\rho}.
\end{equation}

In order to find all synchronization-enforcing mechanisms represented by a single normal Lindblad operator, we proceed by discussing a seemingly reverse problem: the description of the attractor space of a given QMDS and its role in synchronization of states. Characterizing all possible attractor spaces of synchronized two-qubit asymptotic evolutions, we will be able to identify all synchronization-enforcing QMDSs with generator \eqref{LindbladOneL} as those, whose attractor space coincides with one of the asymptotically synchronized systems. According to the theorem \ref{StructureTheorem}, individual elements of the attractor space of a QMDS are determined by equations \eqref{StructureTheoremEq1} and \eqref{StructureTheoremEq3}. Therefore, every possible attractor space corresponds to a particular set of Lindblad operators, each of which generates a QMDS with that attractor space. This link between a Lindblad operator and the ability of the associated QMDS to enforce synchronization on a pair of qubits will be used for their classification.

In order to start, for a given QMDS, the commutation relations \eqref{StructureTheoremEq3} can be used to split the space of two-qubit operators $\bh$ into five subspaces $X_{\omega}$, based on the commutator with the Hamiltonian and a corresponding eigenvalue $\lambda = -i\omega$,

\begin{equation}
\begin{split}
X_{0} = \mathrm{span} \{  \ketbra{00}{00}, \ketbra{01}{01}, \ketbra{01}{10},
\ketbra{10}{01}, \ketbra{10}{10}, \ketbra{11}{11} \}, 
\end{split}
\end{equation}
\begin{equation}
X_{2 \Delta E} = \mathrm{span} \{  \ketbra{00}{11} \},
\end{equation}
\begin{equation}
X_{-2 \Delta E} = \mathrm{span} \{ \ketbra{11}{00} \},
\end{equation}
\begin{equation}
\begin{split}
X_{\Delta E} = \mathrm{span} \{  \ketbra{00}{01}, \ketbra{00}{10}, \ketbra{01}{11}, \ketbra{10}{11} \},
\end{split}
\end{equation}
\begin{equation}
\begin{split}
X_{-\Delta E} = \mathrm{span} \{  \ketbra{01}{00}, \ketbra{10}{00}, \ketbra{11}{01}, \ketbra{11}{10} \}.
\end{split}
\end{equation}

Any attractor associated with a particular eigenvalue $\lambda \in \sigma_{as}(\mathcal{L})$ belongs to the corresponding subspace $X_{i\lambda} \equiv X_{\omega}$ and any element $X \in X_{\omega}$, not necessarily belonging to $\mathrm{Att}(\mathcal{T})$, satisfies $\comm{H}{X} = \omega X$.

The first subspace, $X_0$, represents the stationary part of a possible asymptotic state and as such is irrelevant for synchronization of phases. However, it plays an important role later in addressing the question of complete synchronization.

The next two subspaces, $X_{2\Delta E}$ and $X_{-2\Delta E}$, may contribute only to the asymptotic evolution of the composite system yet do not affect the reduced single-qubit states as any elements $X_1 \in X_{2\Delta E}$ and $X_2 \in X_{-2\Delta E}$ satisfy $\Tr_A X_1 = \Tr_B X_1 = 0$ and $\Tr_A X_2 = \Tr_B X_2 = 0$ respectively.

Finally, the last two subspaces, $X_{\Delta E}$ and $X_{-\Delta E}$, are the only ones corresponding to the non-stationary asymptotic evolution of the reduced single-qubit states. Additionally, the two subspaces are connected by the operation of complex conjugation and, consequently, solving the commutation relations \eqref{StructureTheoremEq1} and \eqref{StructureTheoremEq3} for one of the subspaces provides the solution also for the other one. Indeed, the Lindbladian preserves hermiticity, $\mathcal{L}(\dg{X}) = \mathcal{L}(X)^\dagger$, and thus 
$X$ is an eigenvector of $\mathcal{L}$ with eigenvalue $\lambda$ if and only if $\dg{X}$ is an eigenvector of $\mathcal{L}$ with eigenvalue $\bar{\lambda}$. \\

For this reason, we restrict ourselves to work only with the space $X_{\Delta E}$ and choose to parameterize a general element $X \in X_{\Delta E}$ as

\begin{equation}
\label{parametrisace}
X = \a \ketbra{00}{01} + \b \ketbra{00}{10} + \g \ketbra{01}{11} + \d \ketbra{10}{11}	
\end{equation}
where $\a,\b,\g,\d \in \mathbb{C}$. Then the condition of synchronization \eqref{defSynchEq} reduces to

\begin{equation}
\label{SynchCondX}
\Tr_A X = \Tr_B X,
\end{equation}

which in the chosen parameterization reads

\begin{equation}
\label{SynchCondParameters}
\a + \d = \b + \g.
\end{equation}

In section \ref{section_synchronization}, we defined synchronization only for states, not arbitrary elements of an attractor space. However, the definition extends naturally. Consider a synchronization-enforcing QMDS, whose possible asymptotic states all satisfy \eqref{defSynchEq}. Due to the form of the asymptotic evolution \eqref{AsymptoticDynamics}, the reduced subsystem asymptotic dynamics must obey

\begin{equation}
	\label{SynchCondAttractors}
	\sum_{\substack{\lambda \in \sigma_{as}(\mathcal{L}), i \\ \lambda \ne 0}} e^{\lambda t}\Tr\left\{\left(X^{\lambda, i}\right)^{\dagger}\rho(0)\right\} \left[\Tr_A X_{\lambda , i}- \Tr_B X_{\lambda , i}\right] = 0
\end{equation}

for any two-qubit initial state $\rho(0)$. Since any operator can be expressed as a linear combination of no more than four density operators, the relation \eqref{SynchCondAttractors} is also valid for any operator in place of $\rho(0)$. Thus, it must be satisfied for all attractors $X_{\lambda,i}$ contributing to the asymptotic evolution. Their orthonormal properties imply $\Tr_A X_{\lambda , i}- \Tr_B X_{\lambda , i}=0$ for any $\lambda \neq 0$ from $\sigma_{as}(\mathcal{L})$, a non-trivial condition only for attractors lying in the subspaces $X_{\Delta E}$ and $X_{-\Delta E}$ respectively , i.e. equation \eqref{SynchCondX} with $X \in X_{\Delta E}$. Conversely, at any point in time in the asymptotics, every possible state of the system can be written as a linear combination of the elements of the attractor space \eqref{AsymptoticDynamics}. Therefore, if every attractor satisfies the synchronization condition, so does any asymptotic state.

This is a key observation. The asymptotic state will be synchronized, irrespective of the initial state, if and only if the attractor space is formed by attractors satisfying the condition of synchronization themselves. And it follows from the previous discussion that this requirement is non-trivial only for the subspace $X_{\Delta E}$. \\

In light of theorem \ref{StructureTheorem}, elements of the attractor space of a given QMDS are further determined by the commutation relations \eqref{StructureTheoremEq1}, which in the case of a generator \eqref{LindbladOneL} with a single Lindblad operator read

\begin{equation}
	\label{commutationRelation}
	\comm{L}{X} = \comm{\dg{L}}{X} = 0.
\end{equation}

Hence, the task is to find all possible normal Lindblad operators $L$ such that the solution to the above equation \eqref{commutationRelation} for $X \in X_{\Delta E}$ is non-trivial and satisfies the condition of synchronization \eqref{SynchCondX}. 

Our strategy is to consider consecutively all such possible solutions and for each $X$ find the set of all normal operators $L$ commuting with it. We refer to these sets as {\it sets of commuting Lindblad operators} for simplicity and stress that the contained Lindblad operators generally do not commute with each other. Subsequently, we extract from these sets the operators that enforce the synchronization condition for every $X \in X_{\Delta E}$.
Additionally, since by definition the synchronization condition \eqref{defSynchEq} is also necessary for the stronger condition of complete synchronization \eqref{defCompleteSynchEq}, we make use of the results to further identify all normal Lindblad operators $L$ enforcing complete synchronization. Those are operators $L$ which in addition to \eqref{SynchCondX} enforce $\Tr_A X = \Tr_B X$ also for all attractors $X \in X_0$, as follows directly from the definition \eqref{defCompleteSynchEq}. \\

The same reasoning can be followed for phase-locking and generalized complete synchronization. Replacing the definition of synchronization by that of phase-locking will effectively turn \eqref{SynchCondX} into

\begin{equation}
\label{GenSynchCondX}
\Tr_A X = e^{i\varphi} \Tr_B X,
\end{equation}

for $X \in X_{\Delta E}$, and \eqref{SynchCondParameters} into

\begin{equation}
\label{GenSynchCondParameters}
\a + \d = e^{i\varphi} \left( \b + \g \right),
\end{equation}

where $\varphi \in  [0, 2\pi)$ denotes the delay phase shift. With phase-locking being a straightforward generalization of synchronization, both cases are closely related and will be solved simultaneously. \\

Based on the previous discussion, our analytical approach to identifying synchronization and phase-locking mechanisms can be summarized into few main points:

\begin{itemize}
    \item[1.]  Using the parameterization \eqref{parametrisace}, we consider a general attractor $X \in X_{\Delta E}$ satisfying the phase-locking condition \eqref{GenSynchCondParameters} with a fixed phase-shift $\fii \in [0,2\pi)$. We structure the analysis into several cases, assuming consecutively all possible numbers and combinations of non-zero coefficients $\a.\b,\d,\g$.
    \item[2.] For each attractor $X$, the commutation relations \eqref{commutationRelation} yield the set of all normal operators $L$ commuting with the said attractor. These are Lindblad operators which permit phase-locking, as they admit a nontrivial phase-locking-condition-satisfying solution of \eqref{commutationRelation} for $X \in X_{\Delta E}$, yet not necessarily enforce it. Nonetheless, they are crucial both as an intermediate step and in the discussion of phase-locking with finitely many Lindblad operators.
    \item[3.] From these sets of Lindblad operators those are selected that enforce the phase-locking condition \eqref{GenSynchCondParameters} on the whole subspace $X_{\Delta E}$. This is done by solving \eqref{commutationRelation} yet again, this time for the particular cases of $L$ and a general $X \in X_{\Delta E}$, requiring that all solutions $X$ satisfy \eqref{GenSynchCondParameters}, thereby obtaining additional constrains on $L$ necessary to enforce phase-locking. A natural classification based on the corresponding original attractor $X \in X_{\Delta E}$ arises in the process.
    \item[4.] The operators from the resulting classes of phase-locking-enforcing Lindblad operators and corresponding supersets of operators commuting with the associated attractor can be combined to construct all phase-locking-enforcing Lindbladians with finitely many normal Lindblad operators. We postpone the discussion to section \ref{sec:more_Linblad operators}.
    \item[5.] Repeating the third step for a general $X \in X_{0}$ and a requirement that the solution satisfies $\Tr_A X = \Tr_B X$, Lindblad operators additionally enforcing synchronization of the stationary parts of asymptotic states, i.e. generalized complete synchronization, are identified in section \ref{sec:generalized_complete_synchronization}.
\end{itemize}

This approach constitutes an extensive and lengthy analysis, whose details are presented in \ref{AppendixDerivation}. In the main text we restrict ourselves to merely listing the results, classifying the obtained synchronization and phase-locking mechanisms, and discussing their properties.

\subsection{Catalogue of phase-locking mechanisms with a single Lindblad operator}
\label{sec:overview}

There exist two classes and a one-parameter family of classes of phase-locking-enforcing Lindblad operators, associated each with a particular class-specific attractor $X \in \mathrm{Att}(\mathcal{T}) \cap X_{\Delta E}$ of the generated QMDS $\mathcal{T}$ and a superset of the class formed by all normal Lindblad operators commuting with said attractor.

Before listing them all, let us point out that an overall phase prefactor of a Lindblad operator has no physical meaning due to the form of the generator \eqref{Lindblad}, and therefore we naturally omit them in the derivation in \ref{AppendixDerivation} as well as in the classification presented in this part. \\ 

The first set $L_1$ contains all normal Lindblad operators which commute with the attractor $X = \ketbra{00}{01} + \ee\ketbra{00}{10} \in X_{\Delta E}$ and read
\begin{equation}
	\label{L1}
	\renewcommand\arraystretch{1.2}
	\begin{split}
		L_1 = 
		\begin{pmatrix}
			1 & 0 & 0 & 0 \\
			0 & \frac{1}{\sqrt{2}} & \frac{1}{\sqrt{2}} & 0 \\
			0 & \frac{e^{i\varphi}}{\sqrt{2}} & -\frac{\e}{\sqrt{2}} & 0 \\
			0 & 0 & 0 & 1 \\
		\end{pmatrix}
		\begin{pmatrix}
			c & 0 & 0 & 0 \\
			0 & c & 0 & 0 \\
			0 & 0 & a  & b \\
			0 & 0 & e^{i2k}\bar{b} & a + m e^{ik} \\
		\end{pmatrix}
		\begin{pmatrix}
			1 & 0 & 0 & 0 \\
			0 & \frac{1}{\sqrt{2}} & \frac{\ee}{\sqrt{2}} & 0 \\
			0 & \frac{1}{\sqrt{2}} & -\frac{\ee}{\sqrt{2}} & 0 \\
			0 & 0 & 0 & 1 \\
		\end{pmatrix} \\ = \frac{1}{2}
		\begin{pmatrix}
			2c & 0 & 0 & 0 \\
			0 & c+a & \ee (c-a) & -\sqrt{2}\ee b \\
			0 & \e (c-a) & c+a & \sqrt{2} b \\
			0 & -\sqrt{2} \e e^{2ik} \bar{b} & \sqrt{2} e^{2ik} \bar{b} & 2 (a+me^{ik})
		\end{pmatrix},
	\end{split}
\end{equation}
\nopagebreak
where $a,b \in \mathbb{C}$, $c, k, m \in \mathbb{R}$, and $\varphi \in [0,2\pi)$ denotes the phase shift. The subset of phase-locking-enforcing Lindblad operators, i.e. the class $L_1$, comprises all operators of the form \eqref{L1} which satisfy at least one of the conditions

\begin{equation}
\label{L1ConditionB}
    b \neq 0,
\end{equation}
\begin{equation}
\label{L1ConditionDouble}
    a \neq c \, \land \, m \neq 0.
\end{equation}

The second set of commuting Lindblad operators $L_2$, commuting with the attractor $X = \ketbra{01}{11} + \e\ketbra{10}{11} \in X_{\Delta E}$, comprises operators of the form
\begin{equation}
	\label{L2}
	\renewcommand\arraystretch{1.2}
	\begin{split}
		L_2 =
		\begin{pmatrix}
			1 & 0 & 0 & 0 \\
			0 & \frac{1}{\sqrt{2}} & \frac{1}{\sqrt{2}} & 0 \\
			0 & -\frac{e^{i\varphi}}{\sqrt{2}} & \frac{\e}{\sqrt{2}} & 0 \\
			0 & 0 & 0 & 1 \\
		\end{pmatrix}
		\begin{pmatrix}
			a & b & 0 & 0  \\
			e^{i2k}\bar{b} & a + m e^{ik} & 0 & 0 \\
			0 & 0 & c & 0  \\
			0 & 0 & 0 & c \\
		\end{pmatrix}
		\begin{pmatrix}
			1 & 0 & 0 & 0 \\
			0 & \frac{1}{\sqrt{2}} & -\frac{e^{-i\varphi}}{\sqrt{2}} & 0 \\
			0 & \frac{1}{\sqrt{2}} & \frac{\ee}{\sqrt{2}} & 0 \\
			0 & 0 & 0 & 1 \\	
		\end{pmatrix} \\ = \frac{1}{2}
		\begin{pmatrix}
			2a & \sqrt{2}b & -\sqrt{2}\ee b & 0 \\
			\sqrt{2} e^{2ik} \bar{b} & c + a + me^{ik} & \ee (c-a-me^{ik}) & 0 \\
			-\sqrt{2}\e e^{2ik} \bar{b} & \e (c-a-me^{ik}) & c + a + me^{ik} & 0 \\
			0 & 0 & 0 & 2c \\
		\end{pmatrix},
	\end{split}
\end{equation}
\nopagebreak
where $a,b \in \mathbb{C}$, $c, k, m \in \mathbb{R}$, and $\varphi \in [0,2\pi)$ denotes the phase shift. The corresponding class of phase-locking-enforcing Lindblad operators $L_2$ contains all operators \eqref{L2} for which at least one of the conditions
\begin{equation}
\label{L2ConditionB}
    b \neq 0,
\end{equation}
\begin{equation}
\label{L2ConditionDouble}
    a + me^{ik} \neq c \, \land \, m \neq 0
\end{equation}
holds. Notice the close similarity between the two classes. In fact, there is a one-to-one correspondence between them and also all their studied properties are analogous. This holds also for the corresponding supersets of commuting Lindblad operators.

Additionally, there is an overlap of these two classes. For $b=0$, $m = e^{-ik}(c-a)$ in \eqref{L1} and $b=0$, $c = a$ in \eqref{L2} the operators $L_1$ and $L_2$ both reduce to a Lindblad operator $L_s$ of the form 

\begin{equation}
	\begin{split}
		\label{SWAPclass}
		L_{s} &= j
		\underbrace{\begin{pmatrix}
				1 & 0 & 0 & 0 \\
				0 & 1 & 0 & 0 \\
				0 & 0 & 1 & 0 \\
				0 & 0 & 0 & 1 \\
		\end{pmatrix}}_{I_{4 \times 4}} + \, \, k
		\underbrace{\begin{pmatrix}
				1 & 0 & 0 & 0 \\
				0 & 0 & \ee & 0 \\
				0 & \e & 0 & 0 \\
				0 & 0 & 0 & 1 \\
		\end{pmatrix}}_{SWAP_\fii},
	\end{split}
\end{equation}
\nopagebreak
where $j,k \in \mathbb{C}$, $k \neq 0$, $\fii \in [0,2\pi)$ denotes the phase shift, and $j = \frac{c + a}{2}, k = \frac{c - a}{2}$ in \eqref{L1}. Note that due to the form of the Lindbladian \eqref{Lindblad} and self-adjointness of the $SWAP_\fii$ operator, the identity in \eqref{SWAPclass} can be dropped altogether, for operators $L_s$ differing only in the parameter $j$ generate the same QMDS. As a special case of $L_s$, the overlap includes the well-known SWAP operator, which swaps the states of the involved qubits and has been extensively studied e.g. in the context of quantum consensus in qubit networks \cite{ConsensusGuodong}. The operator $L_s$ constitutes a straightforward generalization thereof. \\

Finally, the family of sets of Lindblad operators $L_{\theta}$, commuting with the attractor $X = e^{i \theta} \ketbra{00}{01} + i \ee e^{i \theta} \tan{\theta}\ketbra{00}{10} + i\ee e^{-i\theta}\tan{\theta}\ketbra{01}{11} - e^{-i\theta}\ketbra{10}{11} \in X_{\Delta E}$, parameterized by $\theta \in (-\frac{\pi}{2},0) \cup (0,\frac{\pi}{2})$, reads
\begin{equation}
\label{Ltheta}
\renewcommand\arraystretch{1.2}
	\begin{split}
    L_{\theta} =
	\begin{pmatrix}
		1 & 0 & 0 & 0 \\
		0 & i\ee \sin{\theta} & \cos{\theta} & 0 \\
		0 & -\cos{\theta} & -i\e \sin{\theta} & 0 \\
		0 & 0 & 0 & 1\\		
	\end{pmatrix}
	\begin{pmatrix}
		a & b & 0 & 0 \\
		e^{i2k}\bar{b} & a + me^{ik} & 0 & 0 \\
		0 & 0 & a & e^{-2i\theta}b \\
		0 & 0 & e^{2i\theta}e^{i2k}\bar{b} & a + me^{ik} \\
	\end{pmatrix} \\
	\begin{pmatrix}
		1 & 0 & 0 & 0 \\
		0 & -i\e \sin{\theta} & -\cos{\theta} & 0 \\
		0 & \cos{\theta} & i\ee \sin{\theta} & 0 \\
		0 & 0 & 0 & 1\\	
	\end{pmatrix},
	\end{split}
\end{equation}
where $a,b \in \mathbb{C}, \, k, m \in \mathbb{R}$, and $\fii \in [0,2\pi)$ denotes the phase shift. The classes of phase-locking-enforcing Lindblad operators $L_\theta$ comprise operators \eqref{Ltheta} additionally satisfying the constraint
\begin{equation}
\label{LthetaCondition}
    b \neq 0.
\end{equation}

Each value of the parameter $\theta$ specifies a unique attractor $X \in X_{\Delta E}$, determining the class of Lindblad operators $L_{\theta}$ and a corresponding set of commuting Lindblad operators within the family.

This completes the list of all generalized synchronization mechanisms given by Lindbladians with exactly one normal Lindblad operator. The results, though, are easily generalized to Lindbladians with an arbitrary finite number of normal Lindblad operators.

\subsection{Synchronization mechanisms with finitely many Lindblad operators}
\label{sec:more_Linblad operators}

According to the structure theorem \ref{StructureTheorem}, every Lindblad operator in the generator imposes a constraint on the attractor space in the form of commutation relations \eqref{StructureTheoremEq1}. Therefore, any generalized-synchronization-enforcing Lindbladian must consist of Lindblad operators which all commute with a common nontrivial subspace of attractors $X \in X_{\Delta E}$ satisfying the generalized synchronization condition \eqref{GenSynchCondX}, and of which at least one or their combination must simultaneously enforce the said condition.

Having considered every possible attractor $X \in X_{\Delta E}$ in correspondence with asymptotic phase-locking, all commuting normal operators were found and linked to a particular attractor. Since the attractors $X \in X_{\Delta E}$ are unique to the classes and corresponding supersets of commuting Lindblad operators in our classification, it follows that operators from distinct sets of commuting Lindblad operators $L_1$, $L_2$ and $L_{\theta}$ cannot be combined in a single Lindbladian. For the purpose of this discussion, operators from the overlap of $L_1$ and $L_2$ are considered to belong to the same set of commuting Lindblad operators. Such a Lindbladian would result in $\mathrm{Att}(\mathcal{T}) \cap X_{\Delta E} = \{ 0 \}$, and consequently a QMDS with only stationary asymptotic single-qubit states.
To have a nontrivial common attractor $X \in X_{\Delta E}$ satisfying \eqref{GenSynchCondX}, all Lindblad operators in the generator \eqref{Lindblad} have to belong to exactly one set of commuting operators $L_1$, $L_2$ or $L_{\theta}$. Additionally, to enforce the synchronization condition \eqref{GenSynchCondX}, either at least one of the Lindblad operators must belong to the corresponding class of phase-locking-enforcing operators, or an equivalent constraint on the attractor space must be enforced by a subset of the Lindblad operators.

Let us discuss the particular classes. The case is simple for operators within $L_{\theta}$. A QMDS enforces generalized synchronization if all its Lindblad operators are of the form \eqref{Ltheta} and at least one of them also satisfies \eqref{LthetaCondition}. Regarding operators from $L_1$ and $L_2$, the situation is a bit more intricate. Consider $L_1$ as an example. There are two conditions, \eqref{L1ConditionB} and \eqref{L1ConditionDouble}, either of which is sufficient for operators from $L_1$ to be phase-locking-enforcing, and the latter is a conjunction of two conditions. The two conditions in \eqref{L1ConditionDouble} correspond to two separate constraints on the attractor space, both enforced also by operators satisfying \eqref{L1ConditionB}. These constraints, however, need not be imposed by a single Lindblad operator. The condition \eqref{L1ConditionDouble} can be split between two. Therefore, a QMDS enforces generalized synchronization if all its Lindblad operators are of the form \eqref{L1}, and at least one of them also satisfies $b \neq 0$, that is \eqref{L1ConditionB}, or a condition equivalent to \eqref{L1ConditionDouble} holds, i.e. at least one of the Lindblad operators satisfies $a \neq 0$ and at least one, whether a different one or the same one, satisfies $m \neq 0$. Similar reasoning can be followed for operators $L_2$ given by \eqref{L2} and conditions \eqref{L2ConditionB}, \eqref{L2ConditionDouble}.

Naturally, the phase-shift parameter $\varphi$ must have the same value for all Lindblad operators in the generator. Note also that the weights of individual operators do not affect the asymptotic evolution and only impact upon the transient stage and convergence rates.

Remarkably, as nearly all of the commuting operators were shown to also enforce generalized synchronization, the generators with finitely many normal Lindblad operators constitute just a straightforward generalization of generators with a single Lindblad operator. The attractor spaces of phase-locking-enforcing generators with finitely many normal Lindblad operators are the same as of those with a single Lindblad operator and therefore also their asymptotic dynamics \eqref{AsymptoticDynamics} coincide. From this point of view, phase-locking-enforcing generators with finitely many normal Lindblad operators bring very little new insight into the studied phenomenon.

\subsection{Generalized complete synchronization with finitely many Lindblad operators}
\label{sec:generalized_complete_synchronization}

Lastly, let us address the question of generalized complete synchronization. In addition to phase-locking of the dynamical parts of single-qubit asymptotic dynamics, it also requires equality of the stationary parts. As implied by the structure of attractor spaces of phase-locking QMDS listed in the overview in section \ref{sec:AttractorSpaces}, generalized complete synchronization is in fact enforced by all phase-locking mechanisms constructed from Lindblad operators from classes $L_1$ and $L_2$. Concerning the classes of operators $L_{\theta}$, generalized complete synchronization is asymptotically reached if and only if $\theta = \pm\frac{\pi}{4}$, in which case \eqref{Ltheta} reduces to 

\begin{equation}
	\label{LalfaComplete}
	\renewcommand\arraystretch{1.2}
	    \begin{split}
    	    L_{\pm\frac{\pi}{4}} = 
    		\begin{pmatrix}
    			1 & 0 & 0 & 0 \\
    			0 &  \pm \frac{i \ee}{\sqrt{2}} &  \frac{1}{\sqrt{2}} & 0 \\
    			0 & - \frac{1}{\sqrt{2}} & \mp \frac{i \e}{\sqrt{2}} & 0 \\
    			0 & 0 & 0 & 1 \\
    		\end{pmatrix}
    		\begin{pmatrix}
    			a & b & 0 & 0 \\
    			e^{2ik}\bar{b} & a+me^{ik} & 0 \vphantom{\frac{\mp i \ee}{\sqrt{2}}} & 0 \\
    			0 & 0 \vphantom{\frac{\mp i \ee}{\sqrt{2}}} & a & \mp ib \\
    			0 & 0 & \pm i e^{2ik}\bar{b} & a+me^{ik} \\
    		\end{pmatrix} \\
    		\begin{pmatrix}
    			1 & 0 & 0 & 0 \\
    			0 &   \mp \frac{i \e}{\sqrt{2}} &  -\frac{1}{\sqrt{2}} & 0 \\
    			0 & \frac{1}{\sqrt{2}} & \pm \frac{ i \ee}{\sqrt{2}} & 0 \\
    			0 & 0 & 0 & 1  \\
    		\end{pmatrix},
    	\end{split}
\end{equation}

where $a, b \in \mathbb{C}$, $\, k, m \in \mathbb{R}$, and $\fii \in [0,2\pi)$ denotes the phase shift. \\

\section{Asymptotic properties of phase-locking-enforcing QMDS and corresponding Lindblad operators}
\label{sectionAsymptoticProperties}

While each of the phase-locking mechanisms yields a pair of asymptotically mutually phase-locked qubits, their asymptotic dynamics as such, local as well as global, differ. This section addresses these differences and studies them from several viewpoints.

\begin{figure}[h]
\centering
\subfloat[Distance between two-qubit states]{\includegraphics[width = 2.6in]{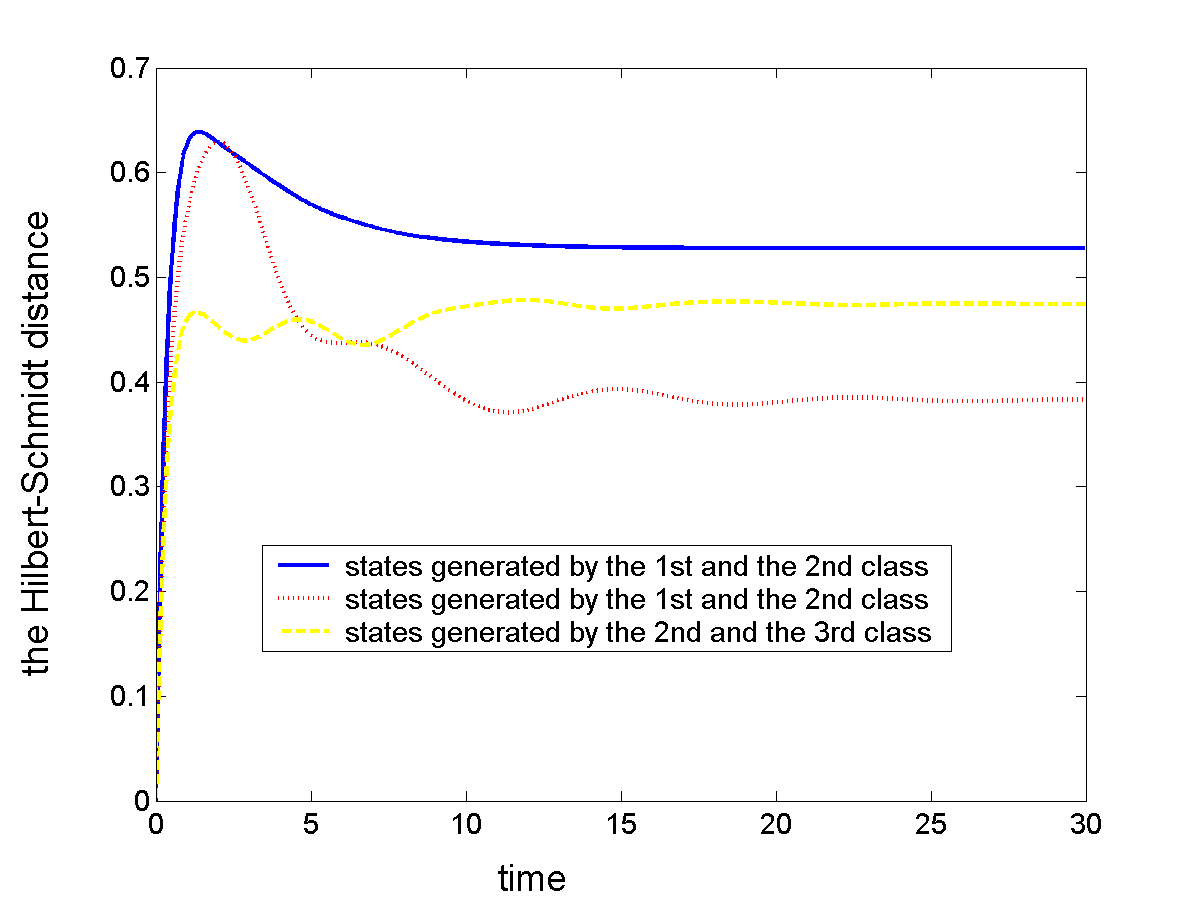}} 
\subfloat[$L_1$ class]{\includegraphics[width = 2.6in]{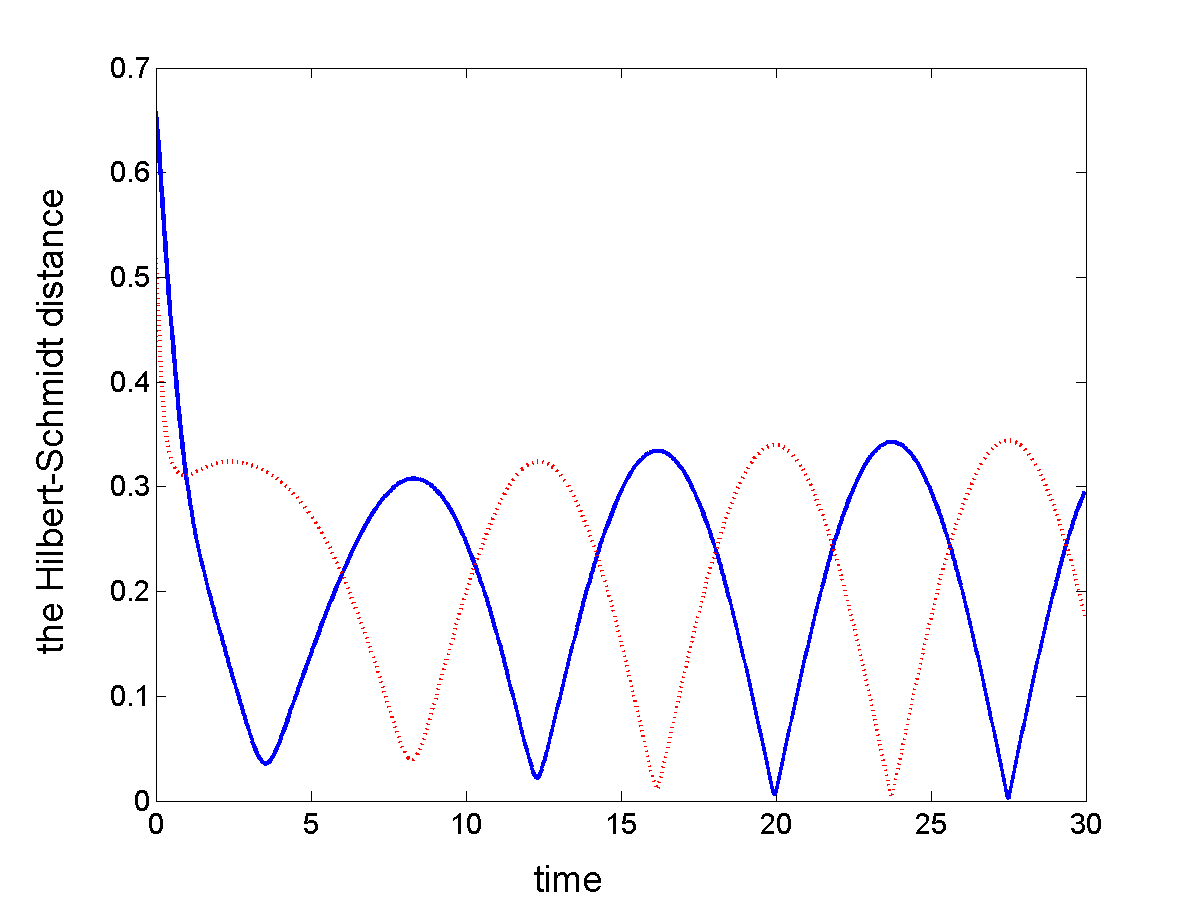}}\\
\subfloat[$L_2$ class]{\includegraphics[width = 2.6in]{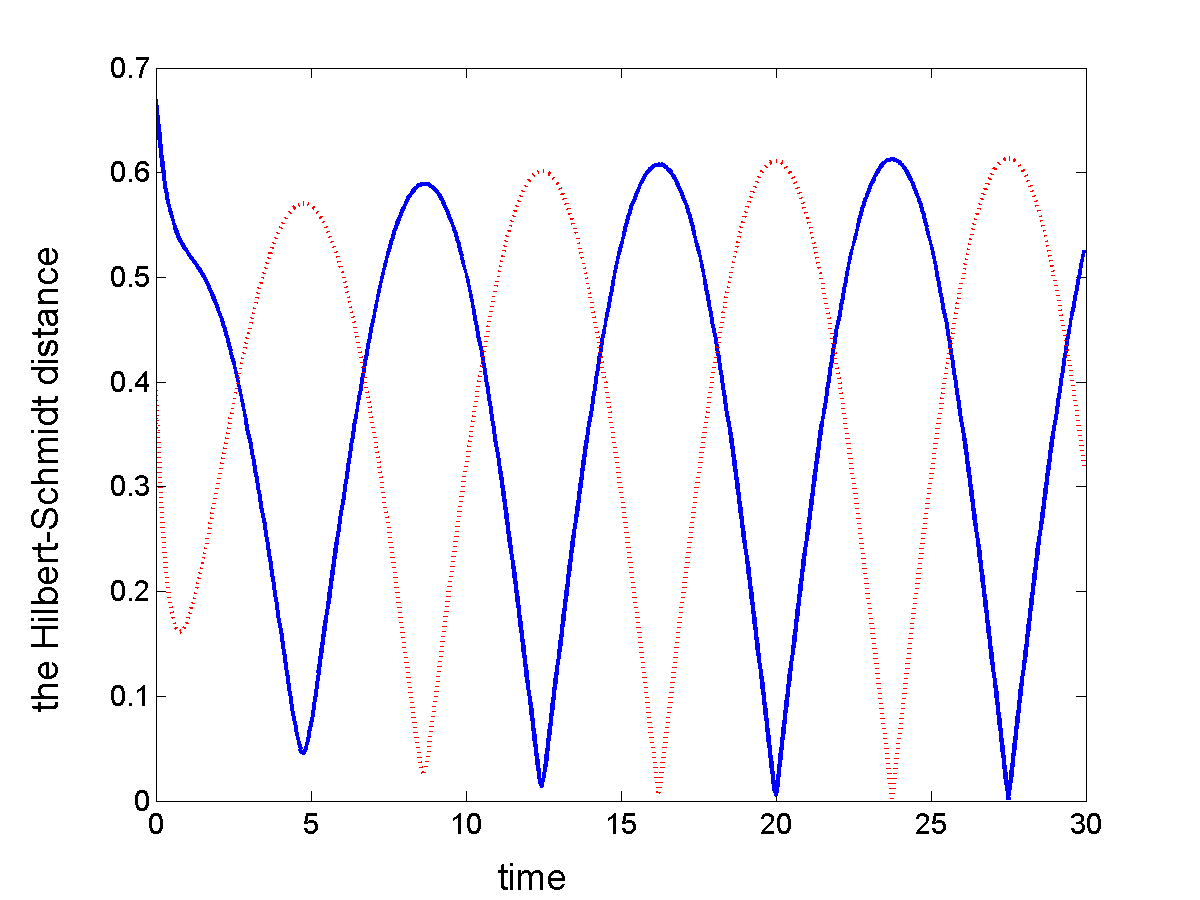}}
\subfloat[$L_{\theta}$ class]{\includegraphics[width = 2.6in]{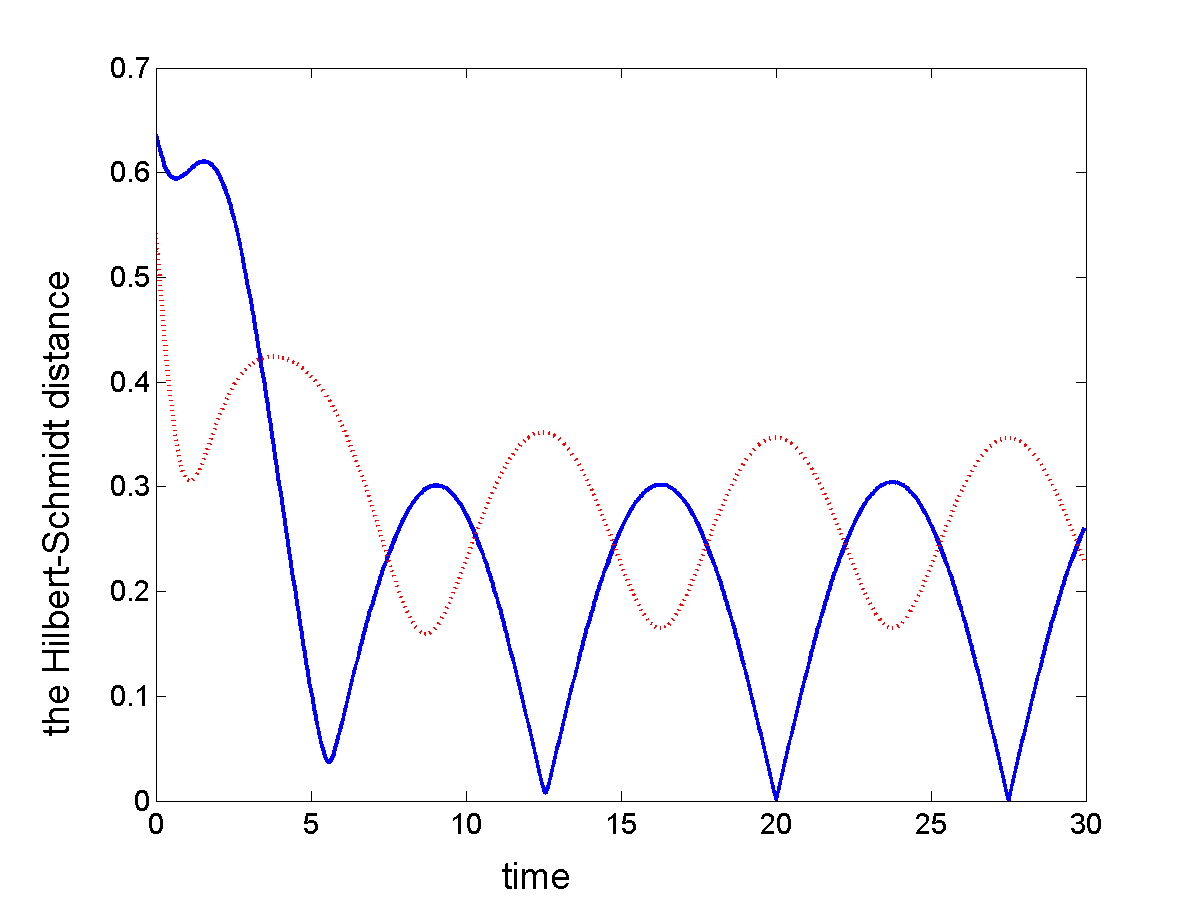}} 
\caption{Comparison of dynamics generated by antisynchronization mechanisms from different classes $L_1$, $L_2$ and $L_{\theta}$, with $\ket{\psi_{in}}$ as the initial state. (a) the Hilbert-Schmidt distance between the 
global two-qubit states corresponding to representatives of individual classes, (b-d) the Hilbert-Schmidt distance between the local single-qubit states corresponding to antisynchronization mechanisms from $L_1$, $L_2$ and $L_{\theta}$ and appropriately chosen local asymptotic states of the first qubit. The blue solid line corresponds to the distance of the first qubit, the red dotted line corresponds to the distance of the second qubit.}
\label{fig:different_mechanisms}
\end{figure}

To give a better idea about the phase-locking processes, let us begin with visualizing them with the use of numerical simulations. Since we investigate mechanisms, each of which leads to a perfect phase-locking in the asymptotic regime, we neither need nor use any measure of the degree of phase-locking. Instead, we directly portray how the individual single-qubit dynamics gradually approach their asymptotics. This is performed by calculating and plotting the time-dependent distance between the evolving single-qubit states and a fixed single-qubit reference state. The Hilbert-Schmidt distance $d(\rho_1,\rho_2)=\sqrt{\Tr\abs{\rho_1-\rho_2}^2}$ between two states $\rho_1$ and $\rho_2$ is used, and except Fig.\ref{fig:different_mechanisms}(a) one of the states is always the reference state, chosen at random as a single-qubit asymptotic state of the first qubit. In the numerical simulations presented throughout this section, we do not specify the choice of the reference state in more detail, as this choice is irrelevant. The asymptotic state being uniquely determined up to a time period of unitary evolution, for sufficiently long time it only shifts the graph in the time axis.

In Fig.\ref{fig:different_mechanisms}, we compare the dynamics generated by representatives of each class of phase-locking mechanisms. 

They are set to antisynchronize two qubits initialized in the state $\ket{\psi_{in}}=\frac{1}{25} \left(3\ket{0}+4\ket{1} \right)\otimes \left(3\ket{0}-4i\ket{1} \right).$ 
The first plot shows the mutual distance between the evolving two-qubit states corresponding to the individual representatives, the other three plots depict the distance between the individual local single-qubit states and a fixed reference local asymptotic state of the first subsystem. In particular, for each representative the initial two-qubit state $\ket{\psi_{in}}$ is evolved towards its asymptotic according to \eqref{LindbladOneL}, both its reduced single-qubit states are expressed and their time-dependent distances from the fixed reference asymptotic state of the first qubit is calculated. Clearly, all three  mechanisms result in antisynchronized yet different dynamics (Fig.\ref{fig:different_mechanisms}(b-d)), which is manifested by the Hilbert-Schmidt distances between the global two-qubit states approaching constant values (Fig.\ref{fig:different_mechanisms}(a)).
Note that because phase-locking mechanisms from the first (Fig.\ref{fig:different_mechanisms}(b), $L_1$) and second (Fig.\ref{fig:different_mechanisms}(c), $L_2$) class also synchronize stationary parts, the respective Hilbert-Schmidt distances from the reference state of both reduced states periodically drop to zero in the asymptotic regime. In general, this does not apply to phase-locking mechanisms within the family of classes $L_{\theta}$ and we observe different oscillations of the first and the second qubit, accounted for by their stationary parts.

The observed differences in behaviour of various phase-locking mechanisms should not be surprising. The requirement to phase-lock qubit dynamics imposes constraints only on a nontrivial subset of attractors from $X_{\Delta E}$, leaving all the other details of the resulting asymptotic dynamics unconstrained. The aim of this section is to explore mutual differences between phase-locking mechanisms from the individual classes and review the efficiency of these processes from various perspectives.

\subsection{Attractors and asymptotic dynamics of synchronization mechanisms}
\label{sec:AttractorSpaces}
This part is devoted to a detailed description of attractor spaces of all phase-locking-enforcing Linbladians presented in section \ref{sec:overview}, their overall asymptotic dynamics and discussion of the actual physical mechanics of how phase-locking is achieved. The attractor spaces, written as linear spans of orthonormal bases of attractors, are listed below. For convenience, the subscript $\omega$ of each attractor denotes the negative imaginary part of its corresponding eigenvalue $\lambda = - i \omega \in \sigma_{as}(\mathcal{L})$ and a subspace $X_{\omega}$ of the attractor space in which the attractor lies \footnote{As such, $\omega$ corresponds to the attractor's commutator with the Hamiltonian, $\comm{H}{X} = \omega X$, and hence to the frequency of its asymptotic unitary evolution.}. In agreement with definition \ref{DefPhaseLocking} and section \ref{section2QSynchronizationMechanisms}, the parameter $\varphi$ denotes the phase shift asymptotically achieved between the reduced single-qubit evolutions. The division as in section \ref{sec:overview} is kept.\\

Denoting $\ket{\psi_1}=\frac{1}{\sqrt{2}}(\ket{01}+e^{i\varphi}\ket{10})$ and $\ket{\psi_2}=\frac{1}{\sqrt{2}}(\ket{01}-e^{i\varphi}\ket{10})$, let $P=\ketbra{00}{00}+\ketbra{\psi_1}{\psi_1}$ and $\tilde{P}=\ketbra{11}{11}+\ketbra{\psi_2}{\psi_2}$ be orthonormal projections satisfying $P + \tilde{P} = I$.

The attractor space $\mbox{Att}(\mathcal{T}_{L_1})$ of a QMDS $\mathcal{T}_{L_1}$ generated by a Lindblad operator from the class $L_1$, given by \eqref{LindbladOneL} and \eqref{L1}, is five-dimensional and reads
\begin{equation}
\label{OverviewX1bnenulove}
\mbox{Att}(\mathcal{T}_{L_1})|_{b \neq 0} = \mbox{span}\left\{\ketbra{00}{00}_0,\ketbra{\psi_1}{\psi_1}_0, \ketbra{00}{\psi_1}_{\Delta E},\ketbra{\psi_1}{00}_{-\Delta E},\left(\frac{1}{\sqrt{2}}\tilde{P}\right)_0
\right\},
\end{equation}
in the case $b \neq 0$ in \eqref{L1}. In this notation the asymptotic two-qubit dynamics generated by this class of phase-locking mechanisms takes a remarkably simple form
\begin{equation}
\label{asymptotics_L1}
    \rho_{as}(t)=e^{-iHt}P\rho(0)P e^{iHt} + \frac{\Tr{\tilde{P}\rho(0)}}{2}\tilde{P}.
\end{equation}
The asymptotic dynamics is obtained as a mixture of a unitarily evolving initial state $\rho(0)$ projected onto the decoherence-free subspace $\mbox{span}\{\ket{00},\ket{{\psi_1}}\}$  \cite{DFS} and a maximally mixed state on its orthogonal complement $\mbox{span}\{\ket{11},\ket{{\psi_2}}\}$.

For $b = 0$, $m \neq e^{-ik}(c-a)$  in \eqref{L1}, the fifth attractor $\frac{1}{\sqrt{2}}\tilde{P}$ splits into two, namely

\begin{eqnarray}
\label{OverviewX1}
\mbox{Att}(\mathcal{T}_{L_1})|_{\substack{b = 0, \\ e^{ik} m \neq c-a}} = \mbox{span}&&\left\{\ketbra{00}{00}_0,\ketbra{\psi_1}{\psi_1}_0, \ketbra{00}{\psi_1}_{\Delta E}, \right. \nonumber \\ &&\left.  \ketbra{\psi_1}{00}_{-\Delta E},\ketbra{\psi_2}{\psi_2}_0,\ketbra{11}{11}_0
\right\},
\end{eqnarray}

and analogously the two-qubit asymptotic dynamics reads 
\begin{equation}
\label{asymptotics_L1_6_attractors}
    \rho_{as}(t)=e^{-iHt}P\rho(0)P e^{iHt} + \langle 11|\rho(0)|11\rangle \ketbra{11}{11} + \langle \psi_2|\rho(0)|\psi_2\rangle \ketbra{\psi_2}{\psi_2}.
\end{equation}
Again, it is a mixture of a unitarily evolving projection of the initial state onto the decoherence-free subspace $\mbox{span}\{\ket{00},\ket{{\psi_1}}\}$ and a completely decohered state on its orthogonal complement $\mbox{span}\{\ket{\psi_2},\ket{11}\}$.

The attractor spaces of QMDS $\mathcal{T}_{L_2}$ generated by Lindblad operators from the class $L_2$ given by \eqref{L2} take an analogous form, solely with exchanged roles of states $\ket{00}$ and $\ket{11}$. Namely, denoting $Q=\ketbra{11}{11}+\ketbra{\psi_1}{\psi_1}$ and $\tilde{Q}=\ketbra{00}{00}+\ketbra{\psi_2}{\psi_2}$ orthonormal projections satisfying $Q + \tilde{Q} = I$, the attractor space reads

\begin{equation}
\label{OverviewX2bnenulove}
\mbox{Att}(\mathcal{T}_{L_2})|_{b \neq 0} = \mbox{span}\left\{\ketbra{11}{11}_0,\ketbra{\psi_1}{\psi_1}_0, \ketbra{11}{\psi_1}_{-\Delta E},\ketbra{\psi_1}{11}_{\Delta E},\left(\frac{1}{\sqrt{2}}\tilde{Q}\right)_0
\right\},
\end{equation}
in the case $b \neq 0$ in \eqref{L2}, respectively

\begin{equation}
\label{OverviewX2}
\mbox{Att}(\mathcal{T}_{L_2})|_{\substack{b = 0, \\ a \neq c}} = \mbox{span}\left\{\ketbra{11}{11}_0,\ketbra{\psi_1}{\psi_1}_0, \ketbra{11}{\psi_1}_{-\Delta E},\ketbra{\psi_1}{11}_{\Delta E},\ketbra{\psi_2}{\psi_2}_0,\ketbra{00}{00}_0
\right\},
\end{equation}

in the case $b \neq 0$, $a \neq c$ in \eqref{L2}. The dynamics preserves the decoherence-free subspace $\mbox{span}\{\ket{11},\ket{{\psi_1}}\}$ and leads to decoherence in its orthogonal complement.

In the remaining case of phase-locking mechanisms from the overlap of classes $L_1$ and $L_2$, i.e. $b=0$, $m = e^{-ik}(c-a)$ in \eqref{L1} and $b=0$, $c = a$ in \eqref{L2} respectively, when the operators reduce to $L_s$ of the form \eqref{SWAPclass}, the attractor space $\mbox{Att}(\mathcal{T}_{L_{s}})$ of a corresponding QMDS $\mathcal{T}_{L_{s}}$ is ten-dimensional and reads
\begin{equation}
\mbox{Att}(\mathcal{T}_{L_s}) = \mbox{Att}(\mathcal{T}_{L_1}) + \mbox{Att}(\mathcal{T}_{L_2}) + \mbox{span}\left\{\ketbra{00}{11}_{2\Delta E}, \ketbra{11}{00}_{-2\Delta E}\right\}.
\end{equation}
Similarly to previous cases, the asymptotic evolution can be written as a mixture of a unitarily evolving projection of the initial state onto a decoherence-free subspace $\mbox{span}\{\ket{00},\ket{{\psi_1}},\ket{11}\}$ and a projection on its orthogonal complement $\ket{\psi_2}$
\begin{equation}
\label{LsAsymptoticDynamics}
    \rho_{as}(t)=e^{-iHt}P_S \rho(0)P_S e^{iHt} + \langle \psi_2|\rho(0)|\psi_2\rangle \ketbra{\psi_2}{\psi_2},
\end{equation}
where $P_S$ denotes the projection onto $\mbox{span}\{\ket{00},\ket{{\psi_1}},\ket{11}\}$.

A common feature of all phase-locking mechanisms within the classes $L_1$ and $L_2$ is that each of them is equipped with a decoherence-free subspace. For example, the dynamics \eqref{LsAsymptoticDynamics} generated by $L_s$ given by \eqref{SWAPclass} gradually destroys coherences between the subspace $\mbox{span}\{\ket{00},\ket{{\psi_1}},\ket{11}\}$ and the state $\ket{\psi_2}$, leaving unperturbed unitary evolution in the subspace $\mbox{span}\{\ket{00},\ket{{\psi_1}},\ket{11}\}$. This shows the special role of the so-called two-qubit phase-locking basis $\{\ket{00},\ket{\psi_1},\ket{\psi_2},\ket{11}\}$ in understanding phase-locking mechanisms of two-level systems. Indeed, a straightforward analysis of pure unitary two-qubit dynamics governed by the system Hamiltonian $H$ reveals that while the reduced single-qubit dynamics of operators $\ketbra{00}{\psi_1}, \ketbra{11}{\psi_1}$ and their adjoint counterparts exhibit a mutual phase delay $\varphi$, the reduced dynamics of operators $\ketbra{00}{\psi_2}, \ketbra{11}{\psi_2}$ and their adjoint counterparts experience a phase delay $\varphi +\pi$. Thus, for any chosen phase shift $\varphi$ there is a two-partite operator decomposition of a general two-qubit density operator such that one part exhibits a phase delay $\varphi$ and the other part a phase delay $\varphi+\pi$ between the contributions to the reduced single-qubit dynamics. Consequently, one possibility of how to enforce mutual phase-locking is to completely destroy contributions either from the $\mbox{span}\{\ketbra{00}{\psi_1}, \ketbra{11}{\psi_1}, \ketbra{\psi_1}{00}, \ketbra{\psi_1}{11}\}$ or from the $\mbox{span}\{\ketbra{00}{\psi_2}, \ketbra{11}{\psi_2}, \ketbra{\psi_2}{00}, \ketbra{\psi_2}{11}\}$, and that is precisely how phase-locking mechanisms within classes $L_1$ and $L_2$ work.

Nonetheless, the attractors corresponding to the classes of phase-locking mechanisms within the $L_{\theta}$ family show that there is another possibility of how to achieve asymptotic phase-locking of two qubits. The four-dimensional attractor space 	$\mbox{Att}(\mathcal{T}_{L_{\theta}})$ reads
\begin{equation}
\label{Xalpha_alternative_1}
\begin{split}
	\mbox{Att}(\mathcal{T}_{L_{\theta}}) = 
	\mbox{span}\left\{
	\frac{1}{2}
	\begin{pmatrix}
	1 & 0 & 0 & 0 \\
	0 & 1 & 0 & 0 \\
	0 & 0 & 1 & 0 \\
	0 & 0 & 0 & 1 \\
	\end{pmatrix}_0,
		\frac{1}{2}
	\begin{pmatrix}
	1 & 0 & 0 & 0 \\
	0 & -\cos{2\theta} & -i\ee \sin{2\theta} & 0 \\
	0 & i\e \sin{2\theta} & \cos{2\theta} & 0 \\
	0 & 0 & 0 & -1 \\
	\end{pmatrix}_0, \right.\\
	\left. \frac{\cos{\theta}}{\sqrt{2}}\begin{pmatrix}
	0 & e^{i\theta} & i \ee \tan{\theta}e^{i\theta} & 0 \\
	0 & 0 & 0 &  i \ee \tan{\theta} e^{-i\theta} \\
	0 & 0 & 0 & -e^{-i\theta} \\
	0 & 0 & 0 & 0 \\
	\end{pmatrix}_{\Delta E}, \right. \\ \left. 
		\frac{\cos{\theta}}{\sqrt{2}}\begin{pmatrix}
	0 & 0 & 0 & 0 \\
	e^{-i\theta} & 0 & 0 & 0 \\
	- i \e \tan{\theta} e^{-i\theta} & 0 & 0 & 0 \\
	0 &  -i \e \tan \theta e^{i\theta} & -e^{i\theta} & 0 \\
	\end{pmatrix}_{-\Delta E}
\right\}.
\end{split}
\end{equation} 
These attractor spaces have a very different structure compared to those of phase-locking mechanisms within classes $L_1$ and $L_2$. There is no coherence preservation between two different sets of states. As only the last two attractors contribute to the non-stationary part of the asymptotic dynamics, it is helpful to express them in the favourable basis $\{\ket{00},\ket{\psi_1},\ket{\psi_2},\ket{11}\}$, where the third attractor takes the form
\begin{equation}
\label{attractor_L_theta}
X_3 = \frac{1}{2}\left[e^{i2\theta}\ketbra{00}{\psi_1}-e^{-i\varphi}e^{-i2\theta}\ketbra{\psi_1}{11} + \ketbra{00}{\psi_2}+e^{-i\varphi}\ketbra{\psi_2}{11}\right],
\end{equation}
and a similar form is taken by the fourth one. It shows that this attractor comprises both operators which, in purely unitary evolution governed by the Hamiltonian $H$, introduce a phase delay $\varphi$ and those which introduce a phase delay $\varphi+\pi$. However, the form of the attractor $X$ is such that the contribution to the reduced single-qubit states from the last two terms vanishes, and the corresponding parts of the attractor contribute only to the overall two-qubit asymptotic dynamics. Thus, it truly results in mutually phase-locked asymptotic single-qubit dynamics with a phase shift $\varphi$.

\subsection{Phase-locking mechanisms and preserved information} 
Dissipative processes inherently entail partial loss of information about the initial state. In this part, we focus on the phase-locking mechanisms' ability to preserve information - memory of these processes. All generalized synchronization mechanisms are irreversible environment-assisted processes, which shrink the set of initial states into a smaller set of asymptotic states. Information preserved during such processes is characterized by the set of constants of motion, which for a QMDS can be constructed from its attractors. Since atractors of a QMDS always come in mutually conjugated pairs, meaning that if $X_{\lambda,i}$ is an attractor associated with eigenvalue $\lambda  \in \sigma_{as}(\mathcal{L})$, $X^{\dagger}_{\lambda,i}=X_{\overline{\lambda},i}$ is an attractor corresponding to eigenvalue $\bar{\lambda}=-\lambda  \in \sigma_{as}(\mathcal{L})$,
a pair of time dependent hermitian observables $C^{(1)}_{\lambda,i}=e^{\lambda t} X_{\lambda,i} + e^{-\lambda t} X_{\lambda,i} ^{\dagger}$ and $C^{(2)}_{\lambda,i}=i(e^{\lambda t} X_{\lambda,i} - e^{-\lambda t} X_{\lambda,i }^{\dagger})$ can be constructed for each pair of conjugated attractors.
It is straightforward to show that these observables are constants of motion, that is their mean values remain constant along all state trajectories or
\begin{equation}
\expval{C(t)}_{\rho(t)} = \Tr\left[C(t)\rho(t)\right] =  \Tr\left[C(t) \mathcal T_t(\rho(0))\right] = c\,(\rho(0)),
\end{equation}
for any state trajectory $\rho(t) = \mathcal T_t(\rho(0))$, i.e. for any input state. Consequently, a QMDS with a $d$-dimensional attractor space is equipped with $d$ linearly independent constants of motion. Note that constants of motion associated with zero eigenvalue are actually integrals of motion. These $d$ constants of motion constitute the information preserved about the two-qubit initial state and for a given initial state the corresponding $d$ real constants uniquely determine the asymptotic state. Roughly speaking, the more attractors, the more information about the initial state preserved and the less disturbed is the free unitary dynamics given by the Hamiltonian $H$.

Memory of the individual phase-locking mechanisms differs significantly, with the dimension of their attractor spaces ranging from four to ten. While the lowest dimension corresponds to phase-locking operators within classes $L_{\theta}$ \eqref{Ltheta}, the greatest to operators from $L_s$ \eqref{SWAPclass}. The latter can in turn be viewed as generating QMDS which achieve generalized synchronization at the expense of the least loss of information. In fact, it has recently been proven \cite{BoundForAttractors} that the number of linearly independent atrractors $d$ of a QMDS on a $D$-dimensional Hilbert space is limited by $d \leq D^2 -2D+2$. For two qubits this gives an upper bound of ten linearly independent attractors, a limit reached by phase-locking mechanisms with Lindblad operators from the class $L_s$, which as a result have the largest possible number of linearly independent constants of motion.

\subsection{Visibility of oscillations of asymptotically synchronized reduced states}

In order to achieve general synchronization it is necessary to sacrifice a certain part of the internal dynamics. It poses a natural question to what extent remain the local dynamics visible and detectable by means of measurement, once phase-locked by one of the mechanisms described in section \ref{section2QSynchronizationMechanisms}.

A free unitary evolution of a single qubit driven by Hamiltonian $H_0$ has in its eigenbasis $\{\ket{0},\ket{1}\}$ a general form 
\begin{equation}
	\label{QubitDensityMatrix}
	\rho(t) = \begin{pmatrix}
		x & y e^{i\Delta Et} \\
		\bar{y} e^{-i\Delta Et} & 1 - x \\
	\end{pmatrix},
\end{equation}
where $y \in \mathbb{C}$, $\Delta E \in \mathbb{R}$, $0 \leq x \leq 1$, and  $\abs{y} \leq  \sqrt{x-x^2}$ from the positivity of $\rho(t)$. In the asymptotics, the evolution is given by \eqref{AsymptoticDynamics} and hence the coefficients $x,y$ are determined by the overlap of the initial state with relevant attractors.

The information about time evolution can be retrieved from the expectation value $\expval{\sigma}(t) = \Tr {\rho(t)\sigma}$ of a suitable local observable $\sigma$. Since any single-qubit observable can be written as a linear combination of $\sigma$ matrices and identity, whereof only $\sigma_x$ and $\sigma_y$ evolve non-trivially in the Heisenberg picture and expectation values of both are proportional to $\abs{y}\cos(Et + \mu)$ for some phase $\mu$, assuming a general qubit state \eqref{QubitDensityMatrix}, it is sufficient to consider, for example, the observable $\sigma_x$. It holds

\begin{equation}
	\label{expvalSigmax}
	\expval{\sigma_x}(t) = 2\abs{y}\cos(\Delta E t + \mu),
\end{equation}
where $\mu \in \mathbb{R}$ accounts for the phase of $y$. Alternatively, we could express the probabilities $p_1 = \Tr{\rho(t)M_1}$, $p_2 = \Tr{\rho(t)M_2}$ of the corresponding projective measurements $M_1 = \frac{1}{2}(\ket{0}+\ket{1})(\bra{0}+\bra{1})$ and $M_2 = \frac{1}{2}(\ket{0}-\ket{1})(\bra{0}-\bra{1})$, which read
\begin{align}
	p_1(t) &= \frac{1}{2}\left\{1 + 2\abs{y}\cos(\Delta E t+\mu)\right\}, \\
	p_2(t) &= \frac{1}{2}\left\{1 - 2\abs{y}\cos(\Delta E t+\mu)\right\},
\end{align}
and consequently
\begin{equation}
\label{SigmaXProbabilitiesDifference}
	p_1(t) - p_2(t) = 2\abs{y}\cos(\Delta E t + \mu) = \expval{\sigma_x}(t).
\end{equation}
Therefore, the time evolution visibility scales with $\abs{y}$, the absolute value of the off-diagonal qubit density matrix element in the eigenbasis of the Hamiltonian. The greater the $\abs{y}$, the bigger the amplitude of local oscillations in \eqref{expvalSigmax} and \eqref{SigmaXProbabilitiesDifference}, and the easier it is to observe the non-trivial evolution of a qubit in state \eqref{QubitDensityMatrix}.

Let us discuss how well the internal single-qubit dynamics is preserved in the asymptotics for the individual generalized synchronization mechanisms. As to classes $L_1$ and $L_2$, the corresponding QMDS were shown to have a decoherence-free subspace. Thus, if the two-qubit evolution is initiated in this subspace or does not involve coherences with its orthogonal complement, it stays unperturbed with original amplitudes of local qubit oscillations. If, on the contrary, the initial state contains coherences between the decoherence-free subspace and its orthogonal complement, the evolution completely destroys these coherences and on the whole decreases the amplitude of single-qubit oscillations. In the extreme case when the projection of the inital state onto the decoherence-free subspace is trivial, the oscillations are fully suppressed. For example, local single-qubit oscillations of the initial pure two-qubit state $\ket{\psi_{in}}= \frac{1}{\sqrt{2}}\left(\ket{\psi_2}+\ket{11} \right)$ are completely destroyed by any phase-locking mechanisms with Lindblad operators from classes $L_1$ and $L_2$. This strong dependence on initial conditions is inevitable. Since for any of the phase-locking QMDS the intersection of the attractor space and the subspace $X_{\Delta E}$ forms a proper subspace of $X_{\Delta E}$, an initial state can always be chosen such that the asymptotic reduced single-qubit states are stationary in spite of presence of single-qubit oscillations in the initial phases of evolution. Nonetheless, regarding classes $L_1$ and $L_2$ a conclusion can be drawn that the visibility of asymptotic local qubit oscillations quantified by the amplitude in \eqref{expvalSigmax} and \eqref{SigmaXProbabilitiesDifference}, i.e. $2|y|$, is generally not limited by the particular phase-locking mechanism and ranges from $0$ to $1$. Note that the enforced phase-shift $\fii$ between single-qubit evolutions plays no role whatsoever with respect to visibility.

A different situation is met with classes $L_{\theta}$. Due to the asymptotic evolution being given by \eqref{AsymptoticDynamics} and the structure of the attractor space \eqref{Xalpha_alternative_1}, the visibility parameter $2|y|$ depends only on the overlap $\Tr\{X_3^{\dagger}\rho(0)\}$ of the initial state $\rho(0)$ with the attractor $X_3$ (\ref{attractor_L_theta}) and the form of the attractor itself. Based on that the visibility parameter $2|y|$ ranges from $0$ to the maximal value $\frac{1}{2}|sin(2\theta)|$. This maximum is again independent of the acquired mutual phase delay $\varphi$, but it does depend on the parameter $\theta$. While for values of $\theta$ close to $0$ or $\pm \pi/2$ the visibility of the individual dynamics is negligible in the asymptotics, for $\theta = \pm \pi/4$ it reaches its maximum $\frac{1}{2}$. It is only a half of the maximum achieved for classes  $L_1$ and $L_2$. The reason is that these phase-locking mechanisms have no decoherence-free subspaces. In this particular case any initial oscillations of local single-qubit dynamics are suppressed at least to a half of their original amplitudes.

Surprisingly, the maxima of visibility are achieved for phase-locking mechanisms which are simultaneously completely synchronizing. From this point of view the generalized complete synchronization results in, rather counterintuitively, better visibility than the less restrictive synchronization or phase-locking.

\subsection{Global symmetry of synchronized states and synchronization mechanisms}
\label{SubsectionSymmetry}

By their very nature, the mechanisms of complete synchronization make two qubits locally indistinguishable. This section addresses the question whether two synchronized qubits also become indistinguishable globally, from the point of view of the composite two-qubit system, and whether or not their symmetry relates to the symmetry of the synchronizing Lindbladian. Indistinguishability of subsystems of a bipartite quantum state requires the global state to be permutationally invariant, i.e. to be invariant with respect to the exchange of qubits. In such a case no measurement can discern the two subsystems.

For a QMDS $\mathcal{T}$ to enforce asymptotic permutation invariance for an arbitrary initial state, all states lying in its attractor space need to be permutation invariant. Following the same reasoning as in section \ref{section2QSynchronizationMechanisms}, this translates to any element of the attractor space being permutation invariant.
Denoting the two-qubit SWAP operator $\Pi = \ketbra{00}{00} + \ketbra{01}{10} + \ketbra{10}{01} + \ketbra{11}{11}$, the invariance is formally written as
\begin{equation}
	\label{PermInvarianceCond}
	X = \Pi \, X \, \Pi
\end{equation}
$\forall X \in \mbox{Att}(\mathcal{T})$. Having fully described the attractor spaces of all synchronization-enforcing QMDS with normal Lindblad operators, we can directly check whether \eqref{PermInvarianceCond} holds.

Naturally, only complete synchronization mechanisms, comprising classes $L_1$, $L_2$ and $L_{\pm \frac{\pi}{4}}$, need to be considered, as permutation invariance \eqref{PermInvarianceCond} evidently implies complete synchronization \eqref{defCompleteSynchEq}. Regarding classes $L_1$ and $L_2$, including their overlap, permutation invariance of the asymptotic state is enforced for all initial conditions. On the other hand, all attractors of QMDS with Lindblad operators from classes $L_{\pm \frac{\pi}{4}}$ but the identity operator violate \eqref{PermInvarianceCond}. Hence, it is guaranteed that no non-trivial asymptotic state is permutation invariant.

It is hereby shown that not all complete synchronization mechanisms, which make two qubits locally indistinguishable, also make them indistinguishable globally. In the following, we explore the qubit-exchange symmetry of the synchronization mechanisms themselves.

A QMDS acts symmetrically on every two-qubit state if its generating Lindbladian is permutation invariant. Due to permutation invariance of the free Hamiltonian and phase ambiguity of Lindblad operators, a Lindbladian \eqref{Lindblad} is permutation invariant if for its every Lindblad operator $L$ there exists $\nu \in \mathbb{R}$ such that
\begin{equation}
	\label{PermInvCondLCorrected}
	 \Pi L \Pi = e^{i\nu} L.
\end{equation}

This condition is easily checked for all operator classes $L_1$, $L_2$ and $L_{\theta}$ given by \eqref{L1},\eqref{L2} and \eqref{Ltheta}. Clearly, we can immediately exclude all phase-locking mechanisms generating a phase shift $\fii \not\in \{ 0,\pi \}$ and incompletely synchronizing operators within classes $L_{\theta}$ for $\theta \neq \pm \frac{\pi}{4}$, and discuss only the synchronization and antisynchronization mechanisms within classes $L_1$, $L_2$ and $L_{\pm \frac{\pi}{4}}$.

Starting with Lidblad operators in $L_{\pm \frac{\pi}{4}}$, these are permutation invariant for neither $\fii = 0$ nor $\fii = \pi$. Interestingly, for both values of $\fii$ the conjugation by $\Pi$ takes operators from $L_{\frac{\pi}{4}}$ to operators from $L_{-\frac{\pi}{4}}$ and vice versa. Moving to $L_1, L_2$, the case of synchronization, $\fii = 0$, operators from both sets are permutation invariant if $b = 0$ in the parameterization \eqref{L1} and \eqref{L2} respectively. This includes but is not limited to the overlap class $L_s$ of the form \eqref{SWAPclass}, containing the SWAP operator itself. Nevertheless, in the case $b \neq 0$ the symmetry transformation \eqref{PermInvCondLCorrected} maps an operator from $L_1$ or $L_2$ back onto an operator in the same class. The transformation merely changes $b$ to $-b$. This gives us another set of symmetrically acting synchronization mechanisms, namely those with Lindblad operators from $L_1$ and $L_2$ where $b \neq 0$ and $a = c = m = 0$. Such operators satisfy $\Pi L_{1(2)} \Pi = - L_{1(2)}$, and thus the corresponding QMDS remain unchanged by this transformation. Finally, the antisynchronizing operators within classes $L_1, L_2$, $\fii = \pi$, are all permutation invariant.

To sum up, only a small part of the generalized synchronization mechanisms uncovered in section \ref{section2QSynchronizationMechanisms} is invariant with respect to the exchange of qubits. A vast majority of them do not treat both qubits equally. There even exist Lindblad operators whereof corresponding QMDS result in permutation invariant asymptotic states for all initial conditions in spite of the Lindblad operators not being permutation invariant themselves. However, for each such an operator a permutation invariant Lindblad operator can be found within the same class. In particular, that is the case of the synchronizing operators in $L_1$ and $L_2$.
On the other hand, there are Lindblad operators, namely the antisynchronizing operators in $L_1, L_2$, which are permutation invariant yet their corresponding QMDS enforce permutation non-invariant asymptotic states.

\subsection{Generalized synchronization and entanglement}
\label{sec:entanglement}
Since phase-locking generally creates mutually correlated subsystems, it is conjectured that phase-locking of quantum systems may witness or may be witnessed by entanglement generated during the process \cite{TS3oscillatorNetworks,BruderTwoSpins}. Exploiting the set of found phase-locking mechanisms, we show that this is not their general feature. In fact, entanglement can be created, destroyed or kept constant during the process of generalized synchronization, depending on the initial state and the particular choice of phase-locking mechanism. In this part, we provide several analytically treatable cases to illustrate the major types of entanglement-related behaviour of our systems of interest.

To quantify the degree of entanglement between two qubits we employ concurrence \cite{ConcurrenceWootters1}, an entanglement monotone given by relation 
\begin{equation}
\label{concurrence}
    C(\rho)=\max\left\{0,\sqrt{\lambda_1}-\sqrt{\lambda_2}-\sqrt{\lambda_3}-\sqrt{\lambda_4} \right\},
\end{equation}
where $\lambda_i$ are eigenvalues of the spin-flipped operator $\rho (\sigma_y \otimes \sigma_y)\rho^{*}(\sigma_y \otimes \sigma_y)$ in descending order, $\sigma_y$ is the Pauli matrix and the star symbol $^{*}$ denotes complex conjugation. For a general normalized pure state written in the phase-locking basis as

\begin{equation}
\label{ConcurrenceDiscussionStatePsi}
\ket{\psi}=x_1\ket{00} + x_2\ket{\psi_1}+x_3\ket{\psi_2}+x_4\ket{11},
\end{equation}

the concurrence takes a much simpler form
\begin{equation}
    \label{concurrence_pure}
    C(\ket{\psi}\bra{\psi})=2|x_1x_4-\frac{e^{i\varphi}}{2}(x_2^2-x_3^2)|.
\end{equation}
Taking into account the asymptotic dynamics of each mechanism discussed in section \ref{sec:AttractorSpaces}, the general expression (\ref{concurrence_pure}) allows us to identify three different initial pure states and phase-locking mechanisms for which all three main types of entanglement behaviour can be observed.

\begin{figure}[t!]
    \centering
    \includegraphics[width=0.7\textwidth]{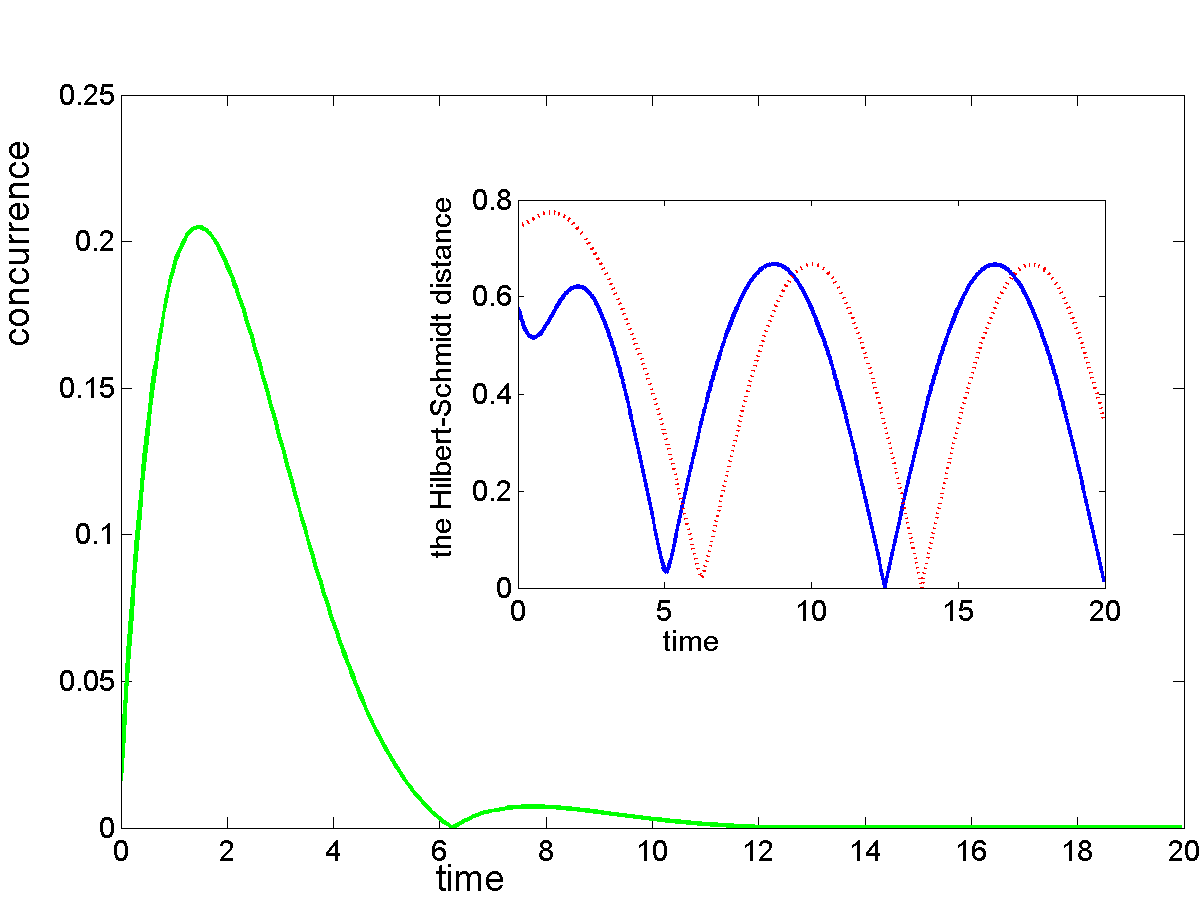}
    \caption{The time evolution of concurrence under the phase-locking mechanism from $L_1$ class with  $\varphi=\pi/3,\Delta E = 4\pi/15,c=1/2,a=3/4+i,b=0,k=0.3,m=0.4$. The initial state is $\ket{\psi}=1/\sqrt{3}\left(\ket{00}+\ket{\psi_1}+\ket{\psi_2}\right)$. The inset plot shows the corresponding mutual phase-locking of qubit dynamics. In particular, the blue-solid (resp. red-dotted) line depicts the Hilbert-Schmidt distance of the actual state of the first (resp. second) qubit from a chosen single-qubit reference asymptotic state.}
    \label{fig:ent_const}
\end{figure}

Let us start with the case when the concurrence of the initial state and the asymptotic phase-locked state is the same, namely zero. This can be easily achieved for an initial pure state $\ket{\psi}$ of the form \eqref{ConcurrenceDiscussionStatePsi} with $x_4=0$ which undergoes evolution under a phase-locking mechanisms from the class $L_1$ with $b = 0$, $m \neq e^{-ik}(c-a)$ in \eqref{L1}. Its asymptotic evolution \eqref{asymptotics_L1_6_attractors} results in a phase-locked state whose concurrence equals

\begin{equation}
    C(\psi|_{x_4=0}) = 2|x_2^2-x_3^2|.
\end{equation}
The situation is shown in Fig.\ref{fig:ent_const}.
The concurrence increases at first and then it falls off towards its original value.

\begin{figure}[h!]
    \centering
    \includegraphics[width=0.7\textwidth]{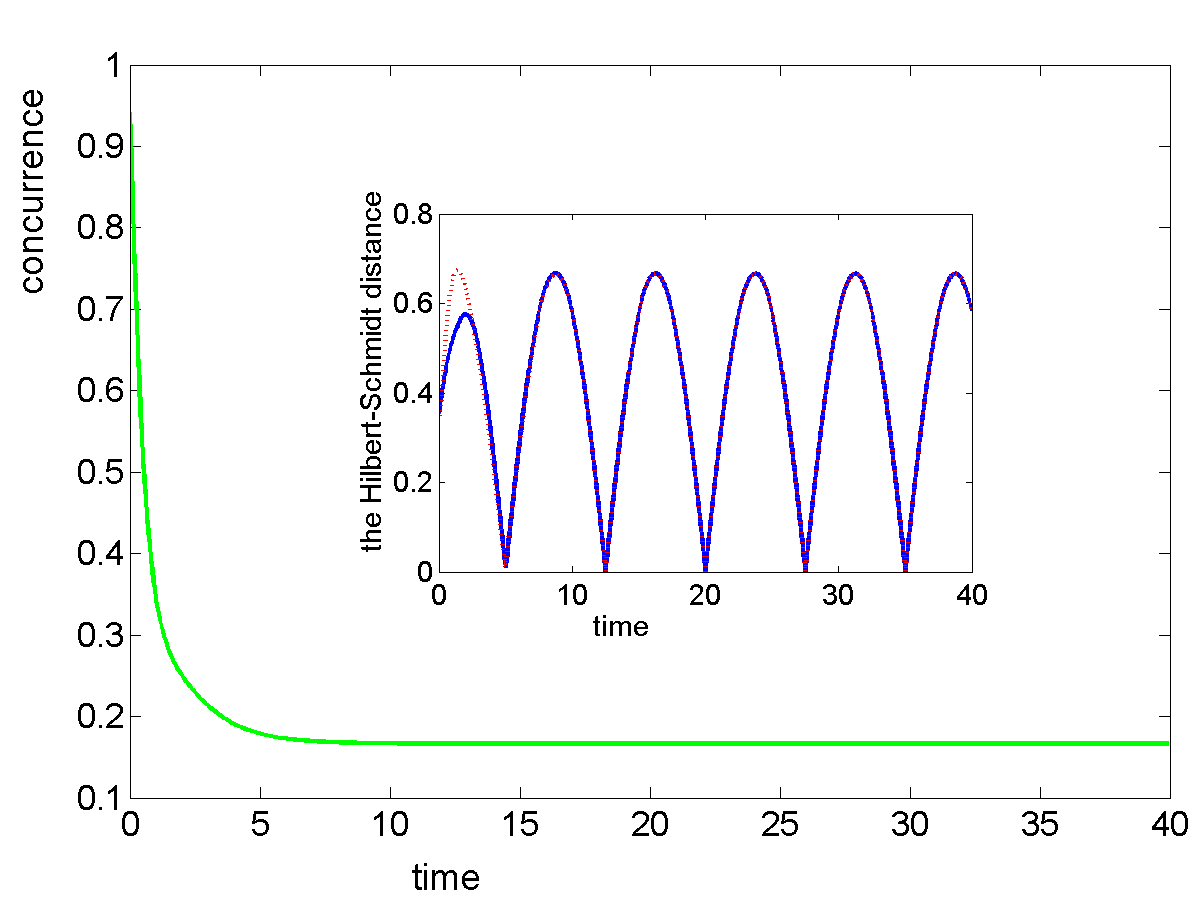}
    \caption{Concurrence behaviour under the evolution generated by phase-locking mechanism from $L_1$ class with parameters $\varphi=0,\Delta E = 4\pi/15,c=1/2,a=3/4+i,b=5/7+i,k=0.3,m=0.4$. The initial state is $\ket{\psi}=1/\sqrt{3}\left(\ket{00}+\exp(i(\varphi/2-\pi/2))\ket{\psi_1}+\ket{11}\right)$. The inset plot shows the corresponding synchronization of qubit dynamics. In particular, the blue-solid (resp. red-dotted) line depicts the Hilbert-Schmidt distance of the actual state of the first (resp. second) qubit from a chosen single-qubit reference asymptotic state.}
    \label{fig:ent_decreasing}
\end{figure}

To observe a net change of concurrence assume an initial state \eqref{ConcurrenceDiscussionStatePsi} where $x_3 = 0$, $x_1=x_4=1/\sqrt{3}$, $x_2=-ie^{i\fii/2}/\sqrt{3}$ and a phase-locking mechanism with Lindblad operators within class $L_1$. Two qubits initialized in a maximally entangled state are driven towards a state with a significantly lower degree of entanglement, quantified by the value of concurrence $C=1/6$ for $b \neq 0$ in \eqref{L1} and $C=1/3$ for $b=0$, independently of all the other parameters. The former case is depicted in Fig.\ref{fig:ent_decreasing}. 
The chosen initial state is also interesting from another reason. Since it is a maximally entangled state, a free evolution would leave it maximally entangled and, consequently, the reduced single-qubit states would remain maximally mixed and locally stationary. The phase-locking mechanism, on the other hand, breaks their entanglement and both qubits establish mutually phase-locked dynamics with significant oscillations.

\begin{figure}[h!]
    \centering
    \includegraphics[width=0.7\textwidth]{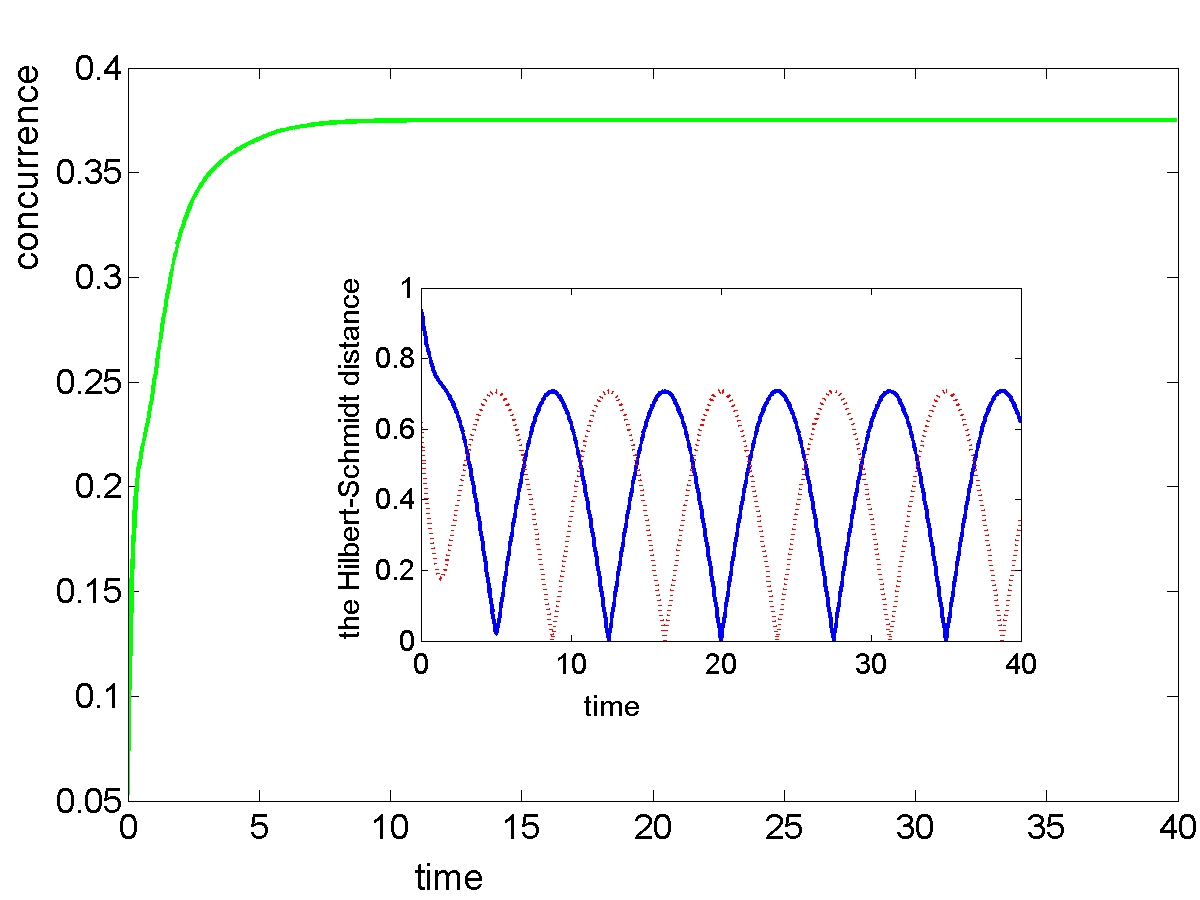}
    \caption{Concurrence behaviour under the evolution generated by phase-locking mechanism from $L_1$ class with parameters $\varphi=\pi,\Delta E = 4\pi/15,c=1/2,a=3/4+i,b=5/7+i,k=0.3,m=0.4$. The initial state is $\ket{\psi}=1/2\left(\ket{00}+\ket{11}\right)+\exp(-i\varphi/2)/\sqrt{2}\ket{\psi_1}$. The inset plot shows the corresponding antisynchronization of qubit dynamics. In particular, the blue-solid (resp. red-dotted) line depicts the Hilbert-Schmidt distance of the actual state of the first (resp. second) qubit from a chosen single-qubit reference asymptotic state.}
    \label{fig:ent_increasing}
\end{figure}

The opposite behaviour is observed for an initial state \eqref{ConcurrenceDiscussionStatePsi} where $x_1=x_4=1/2$ and $x_2=1/\sqrt{2}e^{-i\varphi/2}$. As two-qubits in a factorized pure state gradually become phase-locked their concurrence reaches the value $C=3/8$ for $b \neq 0$ and $C=1/2$ for $b=0$ in \eqref{L1}. This indicates the formation of entanglement between initially factorized systems caused by the environment-assisted interaction. In Fig.\ref{fig:ent_increasing}, we illustrate the situation using an antisynchronization mechanism from the class $L_1$ with $b=5/7+i$.

As can be seen, in none of the scenarios does the behaviour of the degree of entanglement depend on the chosen value of the mutual phase delay $\varphi$. All of the above applies to synchronization as well as to general phase-locking of local dynamics. We point out that phase-locking mechanisms with Lindblad operators from the class $L_2$ have analogous entangling properties as those with operators from the $L_1$ class, and examples of all three presented types of entanglement-related behaviour can be found for phase-locking mechanisms within classes $L_\theta$, too.

We conclude that, despite the general attitude to ascribe to quantum entanglement the ability to witness phase-locking of local dynamics, the analysis discloses a rather inconclusive relation between the two. During a generalized synchronization process entanglement can be destroyed, created or stay constant. If there exists a relationship between phase-locking and entanglement generation, it bears a more intricate form. Nevertheless, types of correlations and the roles they play in the processes of generalized synchronization are a subject worth future studies.

\section{Conclusions}
\label{section_Conclusions}

Ensembles of quantum systems can exhibit a plethora of effects, local or global in their character. Such phenomena are interesting in themselves, find use in technical applications, and naturally shed light on the intricate relations among different quantum features of the analyzed systems.
Synchronization, or more generally phase-locking, is an important collective phenomenon, occurring ubiquitously in nature and studied for centuries. This work concerned phase-locking in a rather general setting, in particular it focused on spontaneously phase-locking continuous quantum Markov processes, which asymptotically compel a mutual phase delay between the individual local dynamics of two qubits, irrespective of the initial state. With an analytical approach based on attractor theory we found all Lindbladians with normal Lindblad operators leading to the desired behaviour. The corresponding phase-locking mechanisms can be categorized as elements of two classes and one family of classes of Lindblad operators. We explained how multiple Lindblad operators can be combined together in a single generator in order to enforce phase-locking, and identified those which, apart from phase-locking, induce equal stationary parts of both qubits, leading to the so-called generalized complete synchronization. 

Equipped with the knowledge of this broad set of phase-locking mechanisms we explored their properties. The complete structure of their attractor spaces was described first, and followed by a detailed analysis of their asymptotic dynamics. It was shown that there are, in fact, only two main types of phase-locking mechanisms. Those of the first type simply destroy coherences outside of a certain decoherence-free subspace containing only mutually phase-locked evolution. The second type comprises a family of classes of phase-locking mechanisms which do not possess any decoherence-free subspace. Instead, they turn any two-qubit initial state into an asymptotic state whose reduced single-qubit density operators contain only the phase-locked components, while contributions to the reduced states with a different mutual phase delay cancel each other. The results reveal the special role of what we call the phase-locking basis, whose analogy we expect to exist in higher-dimensional systems. This paves the way to address the question of synchronization for quantum systems beyond the simplest two-qubit case.

Additional features of phase-locking mechanisms were discussed. In particular the ability to preserve information about the inital state, visibility of the resulting single-qubit oscillations in the asymptotic limit, and permutation symmetry of phase-locking Lindbladians and the way it reflects in the permutation symmetry of the corresponding asymptotic states. Finally, we investigated the role of entanglement in phase-locking processes. Three analytically treatable examples, in which entanglement stays constant, increases and decreases respectively, lead us to the conclusion that if there is a relationship between entanglement and phase-locking it bears a more intricate form than straightforwardly witnessing or being a witness of a successful phase-locking process.

This paper concerns merely two qubits as the simplest nontrivial analytically treatable case, the results and methods, nevertheless, open up the possibility to address the phenomenon of spontaneous synchronization in a broader perspective. We intend to study asymptotic phase-locking processes in many-qubit systems as well as in systems with internal structures more complex than a two-level system and the presented results give hints how to proceed in these situations.

\section{Acknowledgements}
Jaroslav Novotný, Igor Jex and Daniel \v St\v erba acknowledge the financial support from RVO14000 and Centre for Advanced Applied Sciences, Registry No. CZ.02.1.01/0.0/0.0/16 019/0000778, supported by the Operational Programme Research, Development and Education, co-financed by the European Structural and Investment Funds and the state budget of the Czech Republic. Daniel Štěrba acknowledges the support from the Grant Agency of the Czech Technical University in Prague, grant No. SGS22/181/OHK4/3T/14.

\newpage
\bibliographystyle{iopart-num}
\renewcommand\bibname{References}
\bibliography{Bibliography.bib}

\newpage
\appendix

\section{Mechanisms of generalized synchronization}
\label{AppendixDerivation}
In this appendix the technical steps of the analysis of generalized synchronization mechanisms can be found. They follow the approach described in section \ref{section2QSynchronizationMechanisms} and result in the classification of all phase-locking-enforcing normal Lindblad operators and corresponding supersets of commuting Lindblad operators presented in the main text.  Let $X \in X_{\Delta E}$ be an attractor of a QMDS with a generator given by \eqref{LindbladOneL} in the parameterization \eqref{parametrisace} satisfying the condition of generalized synchronization \eqref{GenSynchCondParameters}. Consider consecutively the following possibilities, sorted by the number of nonzero coefficients in \eqref{parametrisace}. \\

Firstly, assume $\d = \g = 0$ and $ \a = e^{i\varphi}\b \neq 0$.
\nopagebreak
In this case, the attractor $X$ reads

\begin{equation}
\label{DynAttL1}
X = \begin{pmatrix}
0 & \a & e^{-i\varphi} \a & 0 \\
0 & 0 & 0 & 0 \\
0 & 0 & 0 & 0 \\
0 & 0 & 0 & 0 \\
\end{pmatrix} \propto \begin{pmatrix}
0 & 1 & e^{-i\varphi} & 0 \\
0 & 0 & 0 & 0 \\
0 & 0 & 0 & 0 \\
0 & 0 & 0 & 0 \\
\end{pmatrix}.
\end{equation}

Noticing that $X = \a \ket{00}\left(\bra{01}+e^{-i\varphi}\bra{10}\right)$ we introduce a new orthonormal basis $(e_1, e_2, e_3, e_4)$ where
\begin{align}
\label{NewBasis1}
e_1 &= \ket{00}, \\
\label{NewBasis2}
e_2 &= \frac{1}{\sqrt{2}}(\ket{01}+e^{i\varphi}\ket{10}), \\
\label{NewBasis3}
e_3 &= \frac{1}{\sqrt{2}}(\ket{01}-e^{i\varphi}\ket{10}), \\
\label{NewBasis4}
e_4 &= \ket{11},
\end{align}

so that the transition matrix
\begin{equation}
\renewcommand\arraystretch{1.2}
\label{TransitionMatrixT1}
T = 
\begin{pmatrix}
1 & 0 & 0 & 0 \\
0 & \frac{1}{\sqrt{2}} & \frac{1}{\sqrt{2}} & 0 \\
0 & \frac{\e}{\sqrt{2}} & \frac{-\e}{\sqrt{2}} & 0 \\
0 & 0 & 0 & 1 \\
\end{pmatrix}
\end{equation}
is unitary and the attractor $X$ in the new basis reads
\begin{equation}
\tilde{X} =
\begin{pmatrix}
0 & \a & 0 & 0 \\
0 & 0 & 0 & 0 \\
0 & 0 & 0 & 0 \\
0 & 0 & 0 & 0 \\
\end{pmatrix} \propto \begin{pmatrix}
0 & 1 & 0 & 0 \\
0 & 0 & 0 & 0 \\
0 & 0 & 0 & 0 \\
0 & 0 & 0 & 0 \\
\end{pmatrix}.
\end{equation}
Using the fact that the commutation relations \eqref{commutationRelation} are invariant with respect to the change of basis we evaluate them directly to obtain $\tilde{L}$, the operator $L$ in the new basis, which can then be transformed back into the original computational basis as $L = T\tilde{L}\ad{T}$. The relation \eqref{commutationRelation} yields a set of equations for the matrix elements of $\tilde{L}$, constraining it to

\begin{equation}
\label{BlockStructure}
\tilde{L} =
\begin{pmatrix}
c\, I_{2\times2} & 0 \\
0 & M \\
\end{pmatrix},
\end{equation}

where $c \in \mathbb{C}$ and $M \in \mathbb{C}^{2x2}$ is an arbitrary complex matrix. It is possible to factor out a phase factor and choose $c \in \mathbb{R}$ instead. The normality condition $\comm{\tilde{L}}{\dg{\tilde{L}}} = 0$, unaffected by the change of basis, can be written in blocks implying that $L$ is normal if and only if the submatrix $M$ is normal. Making use of the parameterization of a general 2x2 normal matrix \eqref{NormalMatrixPar}, whereof derivation is available in \ref{AppendixNormalMatrices}, we arrive at the set of all normal operators $L$ commuting with the attractor \eqref{DynAttL1}. They read
\begin{equation}
\label{resultIIdCandidate}
\renewcommand\arraystretch{1.2}
L = 
\begin{pmatrix}
1 & 0 & 0 & 0 \\
0 & \frac{1}{\sqrt{2}} & \frac{1}{\sqrt{2}} & 0 \\
0 & \frac{e^{i\varphi}}{\sqrt{2}} & \frac{-\e}{\sqrt{2}} & 0 \\
0 & 0 & 0 & 1 \\
\end{pmatrix}
\begin{pmatrix}
c & 0 & 0 & 0 \\
0 & c & 0 & 0 \\
0 & 0 & a  & b \\
0 & 0 & e^{i2k}\bar{b} & a + m e^{ik} \\
\end{pmatrix}
\begin{pmatrix}
1 & 0 & 0 & 0 \\
0 & \frac{1}{\sqrt{2}} & \frac{\ee}{\sqrt{2}} & 0 \\
0 & \frac{1}{\sqrt{2}} & \frac{-\ee}{\sqrt{2}} & 0 \\
0 & 0 & 0 & 1 \\
\end{pmatrix},
\end{equation}

where $a,b \in \mathbb{C}$, $c, k, m \in \mathbb{R}$ and $\varphi \in [0,2\pi)$ is the desired phase shift. \\

To see if the operator $L$ enforces phase-locking via its corresponding QMDS, assume a general attractor $X' \in X_{\Delta E}$ parameterized by $\a',\b',\g',\d' \in \mathbb{C}$ as follows

\begin{equation}
\label{xprime}
X' = \a' \ketbra{00}{01} + \b' \ketbra{00}{10}+ \g' \ketbra{01}{11} + \d' \ketbra{10}{11}.
\end{equation}

In the new basis it can be expressed as $\tilde{X'} = \ad{T}X'T$, which in the matrix form reads

\begin{equation}
\renewcommand\arraystretch{1.1}
\tilde{X'} =  \frac{1}{\sqrt{2}}
\begin{pmatrix}
0 & \a' + e^{i\fii}\b' & \a' - e^{i\fii} \b' & 0 \\
0 & 0 & 0 & \g' + e^{-i\fii}\d' \\
0 & 0 & 0 & \g' - e^{-i\fii}\g' \\
0 & 0 & 0 & 0 \\
\end{pmatrix}.
\end{equation}

As $\tilde{X'}$ is assumed to be an attractor, $\comm{\tilde{L}}{\tilde{X'}}=\comm{\ad{\tilde{L}}}{\tilde{X'}}=0$ holds. Written in the chosen parameterization, it yields the following set of equations

\begin{align}
\label{synch2}
b(\a'-e^{i\fii}\b') &= 0, \\
\label{synch1}
(c-a)(\a'-e^{i\fii}\b') &= 0,\\
\label{synch5}
e^{i2k}\bar{b}(\g'-e^{-i\fii}\d') &= 0, \\
\label{synch6}
me^{ik}(\g'-e^{-i\fii}\d') &=0, \\
\label{synch3}
e^{i2k}\bar{b}(\g'+e^{-i\fii}\d') &=0, \\
\label{synch4}
[c-(a+me^{ik})](\g'+e^{-i\fii}\d') &=0.
\end{align}

The first two constitute constraints on $\a'$ and $\b'$. It follows from \eqref{synch2} and \eqref{synch1} that both $b \neq 0$ and $c \neq a$ implies $\a' = e^{i\fii}\b'$. And since the parameters $\a', \b'$ do not appear in the remaining equations, the requirement

\begin{equation}
\label{cosi1}
b \neq 0 \, \lor \, a \neq c
\end{equation}

is necessary for $L$ to be synchronizing in the generalized sense. However, only the former condition in \eqref{cosi1} is also sufficient. Indeed, for $b \neq 0$ it follows from \eqref{synch5} and \eqref{synch3} that $\g' = \d' = 0$. Consequently, the generalized synchronization condition \eqref{GenSynchCondParameters} holds.

On the other hand, for $b = 0$ the equations \eqref{synch5} and \eqref{synch3} vanish and the parameters $\g', \d'$ are constrained solely by \eqref{synch6} and \eqref{synch4}. In the case of $m = 0$, \eqref{synch6} is trivial and \eqref{synch4} implies $\d' = - e^{i\fii}\g'$, contradicting the generalized synchronization condition. Thus, $m \neq 0$ is needed, in which case the equation \eqref{synch6} yields $\d' = e^{i\fii}\g'$. Depending on the value of $c - a - me^{ik}$, \eqref{synch4} may additionaly compel $\g' = \d' = 0$. In either case, the requirement $m \neq 0$ together with $a \neq c$ is sufficient for $L$ to enforce generalized synchronization. \\

To sum up, a Lindblad operator $L$ given by \eqref{resultIIdCandidate} is synchronizing in the generalized sense if and only if at least one of the conditions
\begin{align}
b &\neq 0, \\
\label{cosi2}
a \neq c \, &\land \, m \neq 0,
\end{align}
is satisfied. Note that excluded from \eqref{resultIIdCandidate} are only the operators which are diagonal in the new basis and such that $\tilde{L}_{11} = \tilde{L}_{22}$, $\tilde{L}_{33} = \tilde{L}_{44}$ or $\tilde{L}_{11} = \tilde{L}_{22} = \tilde{L}_{33}$, an insignificant set of measure zero. We denote this class of synchronization-, respectively phase-locking-enforcing Lindblad operators $L_1$ and write them explicitly in the full form in the main text \eqref{L1}.\\

In exactly the same way, the class of Lindblad operators $L_2$ given by \eqref{L2} is obtained assuming $\a = \b =  0,\, \d = e^{i\varphi} \g$. A step we skip here for the sake of brevity. \\

Secondly, $\a,\b,\g,\d \neq 0$.
The attractor $X$ in the computational basis reads

\begin{equation}
X = \begin{pmatrix}
0 & \a & \b & 0 \\
0 & 0 & 0 & \g \\
0 & 0 & 0 & \d \\
0 & 0 & 0 & 0 
\end{pmatrix},
\end{equation}

where $\a,\b,\g,\d \neq 0$ and $\a + \d = e^{i\fii}(\b + \g)$ holds. Searching for possible submatrices with nonzero determinant in the upper right corner of $X$ it can immediately be seen that $\rank X = 2$. Thus, to simplify evaluation of the commutation relations we introduce a new basis ($e_1,e_2,e_3,e_4$) such that two of the basis vectors span the two-dimensional kernel of $X$ and the other two lie in its orthogonal complement, i.e. $e_1,e_2 \in \Ker X$ and $e_3,e_4 \in (e_1,e_2)^\perp$. Let
\begin{align}
\label{sbasis1}
e_1 &= \ket{00}, \\
e_2 &= \b \ket{01} - \a \ket{10}, \\
e_3 &= \bar{\a}\ket{01} + \bar{\b}\ket{10}, \\
\label{sbasis4}
e_4 &= \ket{11},
\end{align}

and, without loss of generality, the parameters $\a,\b$ are supposed to satisfy a normalization condition

\begin{equation}
\label{normalization}
\abs{\a}^2 + \abs{\b}^2 = 1.
\end{equation}

This is merely a rescalling of the attractor $X$ and as such irrelevant to the result. The normalization ensures that the new basis is orthonormal and the transition matrix

\begin{equation}
\label{unitarybasischange}
T =
\begin{pmatrix}
1 & 0 & 0 & 0 \\
0 & \b & \bar{\a} & 0 \\
0 & -\a & \bar{\b} & 0 \\
0 & 0 & 0 & 1 \\
\end{pmatrix}
\end{equation}

is unitary. Moreover, it holds $Xe_3 = e_1$, with no additional numerical prefactor, which further simplifies the form of $X$ in the new basis. For the remaining basis element $e_4 = \ket{11}$, which comes from the original computational basis, we have $Xe_4 = \g \ket{01} + \d \ket{10}$. Clearly, $Xe_4 \in \mathrm{span}(e_2,e_3)$, a fact that can be used to define two new parameters $s,r \in \mathbb{C}$ via

\begin{equation}
\label{basischange}
Xe_4 = se_2 + re_3,
\end{equation}

to take over the role of the parameters $\d = -s \a + r \bar{\b}$ and $\g = s\b + r\bar{\a}$. This reparameterization helps structurize the disscusion below in simpler terms and holds no physical significance. The attractor $X$ in the new basis reads

\begin{equation}
\tilde{X} =
\begin{pmatrix}
0 & 0 & 1 & 0 \\
0 & 0 & 0 & s \\
0 & 0 & 0 & r \\
0 & 0 & 0 & 0 \\
\end{pmatrix},
\end{equation}

and the partial trace condition of generalized synchronization \eqref{GenSynchCondParameters} takes the form
\begin{equation}
\label{srGenSynch}
(1-s)\a + r\bar{\b} = e^{i\fii}\left[(1+s)\b + r\bar{\a}\right].
\end{equation}

This way we can examine the dependence on the two parameters $s$ and $r$ while the other two, $\a$ and $\b$, keep their role of defining a unitary change of basis \eqref{unitarybasischange}. Again, the result will be of the form $L = T\tilde{L}\ad{T}$.
Rewriting both matrices $\tilde{X}$ and $\tilde{L}$ in a block form 
\begin{equation}
\label{bloky}
\tilde{X} = \begin{pmatrix}
0 & S \\
0 & R \\
\end{pmatrix}, \quad \tilde{L} = \begin{pmatrix}
A & B \\
C & D \\
\end{pmatrix},
\end{equation}
where
\begin{equation}
S = \begin{pmatrix}
1 & 0 \\
0 & s \\
\end{pmatrix}, \quad R = \begin{pmatrix}
0 & r \\
0 & 0 \\
\end{pmatrix},
\end{equation}

introducing matrices $A, B, C, D, S, R \in \mathbb{C}^{2x2}$, the commutation relations \eqref{commutationRelation} imply, among other things, that
\begin{align}
0 = SC = S\ad{B}, \\
0 = RC = R\dg{B}.
\end{align}

Since at least one of the parameters $r,s$ is nonzero it follows
\begin{equation}
B = C = 0,
\end{equation}

and the now block-diagonal form of $\tilde{L}$ further simplifies the commutaion relations into
\begin{align}
\label{cosi3}
SD &= AS, \\
\label{cosi4}
RD &= DR.
\end{align}

The same constraints hold for $\ad{A}, \ad{D}$ in place of $A,D$ as well. Let us consecutively analyze all possible cases for the parameters $s$ and $r$. \\

\text{a)} $s \neq 0\,$:
\nopagebreak

Comparing matrix elements in \eqref{cosi3}, denoting $A = (a_{ij})$, $D = (d_{ij})$, we obtain
\begin{align}
a_{11} &= d_{11}, \\
a_{22} &= d_{22}, \\
d_{12} &= sa_{12}, \\
d_{21} &=  \frac{1}{s}a_{21},
\end{align}
and by doing the same for $\dg{A}, \dg{D}$, taking complex conjugation and comparing with the above we arrive at
\begin{equation}
\label{s}
\bar{s} = \frac{1}{s} \implies \abs{s} = 1.
\end{equation}
If furthermore $r \neq 0$, the relation \eqref{cosi4} implies
\begin{align}
d_{12} = d_{21} = 0 &\implies a_{12} = a_{21} = 0, \\
d_{11}& = d_{22},
\end{align}

and thus the only solutions for $\tilde{L}$ and consequently for $L$ are multiplies of identity, which cannot enforce any form of synchronization. Therefore, we set $r = 0$. Since the matrices $A$ and $D$ are normal due to the block-diagonal shape of $\tilde{L}$ we parameterize $A$, and thus also $D$, using \eqref{NormalMatrixPar}.
From the equation \eqref{srGenSynch} it follows
\begin{equation}
\label{cosi5}
\b = \frac{1-s}{1+s} e^{-i\fii} \a,
\end{equation}
respectively
\begin{equation}
\label{sss}
s = \frac{\a - e^{i\fii}\b}{\a + e^{i\fii}\b}.
\end{equation}
The fact that $\abs{s} = 1$ from \eqref{s} further implies
\begin{equation}
\label{abphasedifference}
    \arg{\b} = \arg{\a} - \fii + \frac{\pi}{2} + k\pi, \quad k \in \mathbb{Z},
\end{equation}
i.e. the phase difference between $\a$ and $\b$ is given by a multiplication factor of $\pm i e^{-i\fii}$.

Note that the seemingly problematic cases $\a = e^{i\fii}\b$, contradicting $s \neq 0$, and $\a = -e^{i\fii}\b$, for which the equation \eqref{sss} is not defined, are excluded as a consequence of \eqref{srGenSynch}. This reflects the fact that setting $s = r = 0$ is equivalent to $e_4 \in \Ker X$ and $\d = \g = 0$, the situation of only two non-zero parameters already discussed above. Analogously, the cases of $s = \pm 1$, apparently relevant to \eqref{cosi5}, correspond to $\b = 0$ and $\a = 0$ respectively, and as such are naturally excluded here. \\

Together with the normalization condition \eqref{normalization}, equations \eqref{abphasedifference} and \eqref{sss} show that the choice of the parameter $\a$ determines two pairs $(\b, s)$, the two possibilities stemming from the two possible phase differences between $\a$ and $\b$. Importantly, they are non-equivalent in the sense that they correspond each to a different attractor
\begin{equation}
\label{DynAttLalpha}
X = \begin{pmatrix}
0 & \a & \b & 0 \\
0 & 0 & 0 & s\b \\
0 & 0 & 0 & -s\a \\
0 & 0 & 0 & 0
\end{pmatrix},
\end{equation}
and set of Lindblad operators 
\begin{equation}
\label{resultIVa}
L =
\begin{pmatrix}
1 & 0 & 0 & 0 \\
0 & \b & \bar{\a} & 0 \\
0 & -\a & \bar{\b} & 0 \\
0 & 0 & 0 & 1\\		
\end{pmatrix}
\begin{pmatrix}
a & b & 0 & 0 \\
e^{i2k}\bar{b} & a + me^{ik} & 0 & 0 \\
0 & 0 & a & sb \\
0 & 0 & \bar{s}e^{i2k}\bar{b} & a + me^{ik} \\
\end{pmatrix}
\begin{pmatrix}
1 & 0 & 0 & 0 \\
0 & \bar{\b} & -\bar{\a} & 0 \\
0 & \a & \b & 0 \\
0 & 0 & 0 & 1\\	
\end{pmatrix},
\end{equation}
where $a, b, \in \mathbb{C}$, $\, k, m \in \mathbb{R}$, $\a \in \mathbb{C}$, $0 < \abs{\a} < 1$,
\begin{align}
\label{abPhaseDiff}
\b &= \pm ie^{-i\fii} \frac{\a}{\abs{\a}} \sqrt{1-\abs{\a}^2}, \\
\label{sExplicit}
s &= \frac{\abs{\a} \mp i \sqrt{1-\abs{\a}^2}}{\abs{\a} \pm i \sqrt{1-\abs{\a}^2}}.
\end{align}
The matrices in \eqref{resultIVa} can be multiplied to reveal that the phase of $\a$ can be included in the parameter $b$, whether $b \neq 0$ or not, without affecting the attractor $X$, as can be seen e.g. from the fact that the attractor $X$ itself is proportional to $\a$. We can therefore choose $\a \in \mathbb{R}$, removing a redundancy in the description.

Finally, the expressions above can be simplified by substituting $\cos{\theta} = \a$ and letting $\theta \in (-\frac{\pi}{2},0) \cup (0,\frac{\pi}{2})$ in order to account for the two possible signs in \eqref{abPhaseDiff}, \eqref{sExplicit}, exploiting opposite parities of the sine and cosine functions. This results in
\begin{equation}
\label{alpha_theta}
\a = \cos{\theta},  
\end{equation}
\begin{equation}
\label{beta_theta}
\b = i \ee \sin{\theta},    
\end{equation}
\begin{equation}
\label{s_theta}
s = e^{-2i\theta},
\end{equation}
and a one-parameter family of classes of Lindblad operators $L_\theta$ of the form \eqref{Ltheta}.\\

Note that the attractor $X$ \eqref{DynAttLalpha} associated with $L \equiv L_\theta$ \eqref{resultIVa} is entirely determined by a single parameter $\theta$. That is why we select the parameter $\theta$ to parameterize the family of classes of Lindblad operators, as opposed to the rest of the parameters which will again specify individual operators within these classes. This way we always have a single class of operators corresponding to a particular attractor $X \in X_{\Delta E}$. \\

In order to determine whether the operators $L$ of the form \eqref{resultIVa} truly enforce synchronization or phase-locking we once again parameterize $X' \in X_{\Delta E}$ as in \eqref{xprime}. In the new basis $\tilde{X'}$ reads

\begin{equation}
\tilde{X'} = 
\begin{pmatrix}
0 & \b\a' - \a\b' & \bar{\a}\a' + \bar{\b}\b' & 0 \\
0 & 0 & 0 & \bar{\b}\g' - \bar{\a}\d' \\
0 & 0 & 0 & \a\g' + \b\d' \\
0 & 0 & 0 & 0 \\
\end{pmatrix}.
\end{equation}

The commutation relations \eqref{commutationRelation}, which now take the form $\comm{\tilde{L}}{\tilde{X'}}=\comm{\dg{\tilde{L}}}{\tilde{X'}}=0$, yield the following set of equations
\begin{align}
\label{synchs1}
0 &= e^{i2k}\bar{b}(\b\a'-\a\b'),\\
\label{synchs2}
0 &= me^{ik}(\b\a'-\a\b'),\\
\label{synchs3}
b(\bar{\b}\g' - \bar{\a}\d') &= sb(\bar{\a}\a'+\bar{\b}\b'),\\
\label{synchs4}
e^{i2k}\bar{b}(\bar{\a}\a'+\bar{\b}\b') &= \bar{s}e^{i2k}\bar{b}(\bar{\b}\g-\bar{\a}\d'),\\
\label{synchs5}
0 &= \bar{s}e^{i2k}\bar{b}(\a\g'+\b\d'),\\
\label{synchs6}
0 &= me^{ik}(\a\g'+\b\d'),
\end{align}
and analogously for $\dg{\tilde{L}}$, resulting in the same set of constraints. Let us first assume the case $b \neq 0$. It follows from \eqref{synchs1} that 
\begin{equation}
\label{synchsres1}
\b' = \frac{\b}{\a}\a',
\end{equation}
and from \eqref{synchs5} that
\begin{equation}
\label{synchsres2}
\g' = -\frac{\b}{\a}\d'.
\end{equation}
Inserting these results into either \eqref{synchs3} or \eqref{synchs4}, multiplying by $\a$ and utilizing the imposed normalization \eqref{normalization} yields
\begin{equation}
\label{cosi6}
\d' = -s\a'.
\end{equation}
The fact that the generalized synchronization condition $\a'+\d'=e^{i\fii}(\b'+\g')$ holds follows, using the relation \eqref{cosi5},
\begin{equation}
\begin{split}
e^{i\fii}(\b'+\g') &\overset{\eqref{synchsres1}\eqref{synchsres2}}{=} e^{i\fii}\frac{\b}{\a}\left(\a' - \d'\right) \overset{\eqref{cosi6}}{=} e^{i\fii}\frac{\b}{\a}\a'(1+s) \\ &\overset{\eqref{cosi5}}{=} \a'(1-s) \overset{\eqref{cosi6}}{=} \a' + \d'
\end{split}	
\end{equation}

This shows that $b \neq 0$ is a sufficient condition for $L$ to enforce phase-locking. \\

On the other hand, consider the case $b = 0$. The equations \eqref{synchs1},\eqref{synchs3},\eqref{synchs4} and \eqref{synchs5} become trivial. If additionally $m = 0$,  the conditions \eqref{synchs2} and \eqref{synchs6} vanish as well, the operator $L$ is a multiple of identity and as such does not enforce any synchronization or phase-locking. Assume therefore $m \neq 0$. The equations \eqref{synchs2} and \eqref{synchs6} are the only non-trivial remaining constraints on the attractor $X'$ stemming from the commutation relations and they retrieve the results \eqref{synchsres1} and \eqref{synchsres2}. The relation \eqref{cosi6} is not enforced in this case and the claim is that, consequently, the generalized synchronization condition does not hold. As a counterexample, let $\a' = \d' = \frac{\a}{\b}\b' = - \frac{\a}{\b}\g' \neq 0$. This is an attractor $X'$ commuting with $L$ and yet not satisfying the generalized synchronization condition as
\begin{equation}
\a'+\d' = 2\a' \neq 0
\end{equation}
does not equal
\begin{equation}
e^{i\fii}(\b' + \g') = e^{i\fii}\frac{\b}{\a}(\a'-\a') = 0.
\end{equation}
This proves that the condition $b \neq 0$ is also necessary.\\

\text{b)} $r \neq 0\,$:

We have already demonstrated that if both $s$ and $r$ are non-zero, the only operators commuting with such attractors $X$ are multiples of identity. Therefore, we assume $s=0$ further on. The relations $SD = AS$ and $RD = DR$ imply
\begin{align}
\a_{11} = d_{11}& = d_{22}, \\
a_{12} = a_{21} = d_{12}& = d_{21} = 0,
\end{align}
and $L$ simplifies into
\begin{equation}
\label{rrr}
L = 		
\begin{pmatrix}
1 & 0 & 0 & 0 \\
0 & \b & \bar{\a} & 0 \\
0 & -\a & \bar{\b} & 0 \\
0 & 0 & 0 & 1\\		
\end{pmatrix}	
\begin{pmatrix}
a & 0 & 0 & 0 \\
0 & b & 0 & 0 \\
0 & 0 & a & 0 \\
0 & 0 & 0 & a \\
\end{pmatrix}
\begin{pmatrix}
1 & 0 & 0 & 0 \\
0 & \bar{\b} & -\bar{\a} & 0 \\
0 & \a & \b & 0 \\
0 & 0 & 0 & 1\\	
\end{pmatrix},
\end{equation}
where $a, b, \a, \b \in \mathbb{C}$, $\abs{\a}^2+\abs{\b}^2=1$. The parameter $r$ is given by
\begin{equation}
r = \frac{\a - e^{i\fii}\b}{e^{i\fii}\bar{\a}-\bar{\b}},
\end{equation}
for $\a \neq e^{i\fii}\b$, and $r$ can be arbitrary for $\a = \e \b$. Due to \eqref{basischange}, $\d = r\bar{\b}$ and $\g = r\bar{\a}$, so that
\begin{equation}
\label{rrratractor}
X = \begin{pmatrix}
0 & \a & \b & 0 \\
0 & 0 & 0 & r \bar{\a} \\
0 & 0 & 0 & r \bar{\b} \\
0 & 0 & 0 & 0
\end{pmatrix}.
\end{equation}

In the case $\a = e^{i\fii}\b$, the parameter $r$ does not affect the operator $L$ or the transition matrix $T$. In fact, $r$ being a free parameter corresponds to $X$ \eqref{rrratractor} splitting into two independent attractors $\ketbra{00}{01} + \e \ketbra{00}{10}$ and $\ketbra{01}{11} + \ee \ketbra{10}{11}$. The operator $L$ reduces to a particular case of $L_2$ \eqref{L2}.

In the case $\a \neq e^{i\fii}\b$, the assumed attractor \eqref{rrratractor} is only fully determined by all three parameters $\a, \b$ and $\fii$. Since the operator $L$ given by \eqref{rrr} does not constrain the phase shift $\fii$ in any way, it does not lead to a synchronizing or phase-locking map.  \\

\text{c)} $s = r = 0\,$:

By the definition of $s$ and $r$ \eqref{basischange}, this is equivalent to the already described case $e_4 \in \Ker X$ and $\d = \g = 0$.\\

The above discussed solutions constitute the full set of normal Lindblad operators capable of enforcing generalized synchronization of a two-qubit system governed by the master equation \eqref{LindbladOneL}. Other possible configurations of parameters $\a,\b,\g,\d$ satisfying the generalized synchronization condition \eqref{GenSynchCondParameters} lead only to trivial solutions to the commutation relations \eqref{commutationRelation}, resulting in phase-locking non-enforcing QMDS.

\newpage
\section{Parameterization of normal matrices}
\label{AppendixNormalMatrices}

Assume a matrix $M \in \mathbb{C}^{2x2}$ parameterized by $a,b,c,d \in \mathbb{C}$,

\begin{equation}
M = \begin{pmatrix}
a & b \\
c & d \\
\end{pmatrix},
\end{equation}

satisfying

\begin{equation}
\comm{M}{\dg{M}} = 0.
\end{equation}

Evaluating this commutation relation element-wise yields the following set of equations

\begin{equation}
\label{AppAprvni}	\abs{b}^2 = \abs{c}^2,
\end{equation}
\begin{equation}
\label{AppAdruha}	a\bar{c} + b\bar{d} = \bar{a}b + \bar{c}d.
\end{equation}

It follows from \eqref{AppAprvni} that we can rewrite $c$ as

\begin{equation}
\label{AppAParB}
c = e^{i2k}\bar{b},
\end{equation}

where $k \in \mathbb{R}$. Inserting that into \eqref{AppAdruha} and rearranging the terms gives

\begin{equation}
\label{AppActvrta}
\underbrace{(d-a)}_{me^{il}}e^{-i2k}b = \underbrace{(\bar{d}-\bar{a})}_{me^{-il}}b,
\end{equation}

denoting $d-a = me^{il}$ with $m,l \in \mathbb{R}$. This equation is trivially satisfied for $b=0$, implying $c=0$ and leaving $a,d$ arbitrary, or for $a = d$, i.e. $m = 0$, leaving $b, k$ arbitrary. In other cases we can divide \eqref{AppActvrta} by its right-hand side. The result reads

\begin{equation}
\label{AppAPomParL}
e^{i2l} e^{-i2k} = 1,
\end{equation}

implying $l = k$. We can drop the other solutions $l + n\pi,\, n\in\mathbb{Z}$, which only introduce unnecessary redundance. Hence

\begin{equation}
\label{AppAParD}
d = a + m e^{ik}.
\end{equation}

Put together, an arbitrary normal matrix $M \in \mathbb{C}^{2\times 2}$ can be parameterized as

\begin{equation}
\label{NormalMatrixPar}
M = \begin{pmatrix}
a  & b \vspace{3pt}\\
e^{i2k}\bar{b} & a + m e^{ik} \\
\end{pmatrix},
\end{equation}
where $a,b \in \mathbb{C}, \, k,m \in \mathbb{R}$.

\end{document}